\documentclass{aims2}
\usepackage{amsmath}
\usepackage{paralist}
\usepackage{graphicx}  
\usepackage{epstopdf}
   
\usepackage[utf8]{inputenc}
   
\usepackage{amssymb}
\usepackage[numbers]{natbib}
\usepackage[colorlinks,citecolor=blue,urlcolor=blue,filecolor=blue]{hyperref}
\usepackage{bm}
\usepackage{soul}
\usepackage{pifont}
\usepackage{xfrac}
\usepackage{bbm} 
\usepackage{bbold} 
\usepackage{array}
\newcolumntype{M}{>{\centering\arraybackslash}m{11.3em}}
\newcolumntype{Q}{>{\centering\arraybackslash}m{12em}}
\usepackage{booktabs}
\usepackage{lipsum}
\usepackage{everypage}
\usepackage{afterpage}
\usepackage{ifthen}
\usepackage{caption}
\usepackage[figuresright]{rotating} 
\usepackage[pagewise]{lineno}
   
\hypersetup{urlcolor=blue, citecolor=red}

\def\ppages{}
\def\DOI{}

\theoremstyle{definition}

\usepackage[boxed]{algorithm2e}
\renewcommand{\figurename}{Fig.}

\usepackage{tikz}
\usetikzlibrary{arrows,calc}
\tikzset{
link/.style={->},
line/.style={thick},
trans/.style={thick,->,shorten >=0.2},
}

\newcommand{\indep}{\mathrel{\text{\scalebox{1.07}{$\mkern-2mu\perp\mkern-10mu\perp\mkern-2mu$}}}}
\newcommand{\notindep}{\mathrel{\text{\scalebox{1.07}{$\not\mkern-2mu\perp\mkern-10mu\perp\mkern-2mu$}}}}

\AddEverypageHook{
  \ifthenelse{\value{page}=8}{
    \afterpage{
\begin{figure}[t!]
\vspace{0.5cm}
{
\setlength\tabcolsep{1pt}
\begin{tabular}{m{0.7em}QMM}
\textbf{1}& \small{$X,Y,Z \sim \mathcal{U}(0,1)$}
&\includegraphics[width=12em]{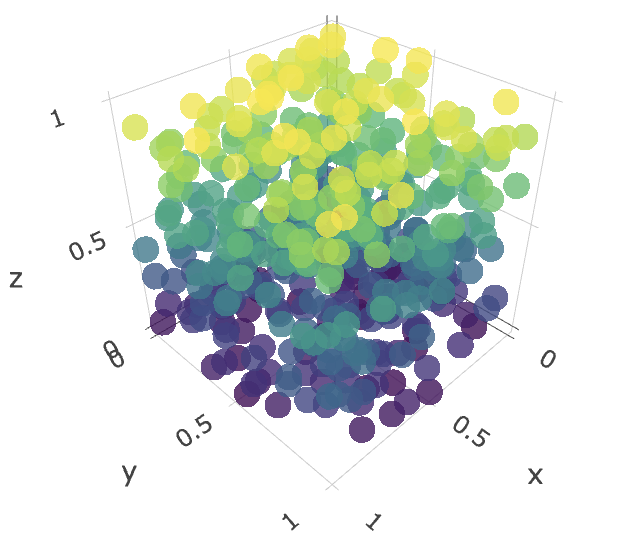} &  \includegraphics[width=11em]{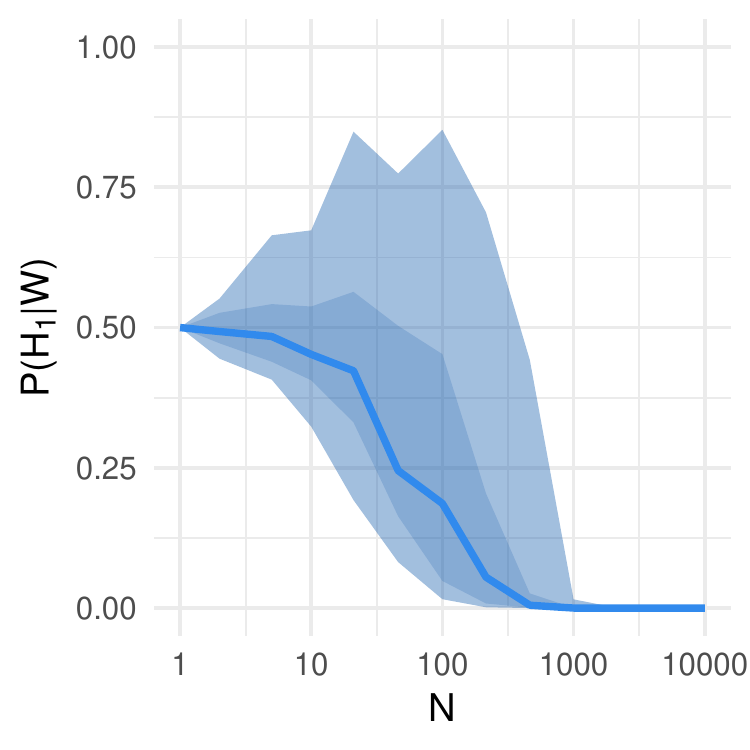} \\
\textbf{2}& \vspace{-3em} \small{
$\begin{array}{c} U,V,Z \sim \mathcal{U}(0,1) \\
	X = 0.5(U + Z) \\ Y = 0.5(V + Z)\end{array}$ }
&\includegraphics[width=12em]{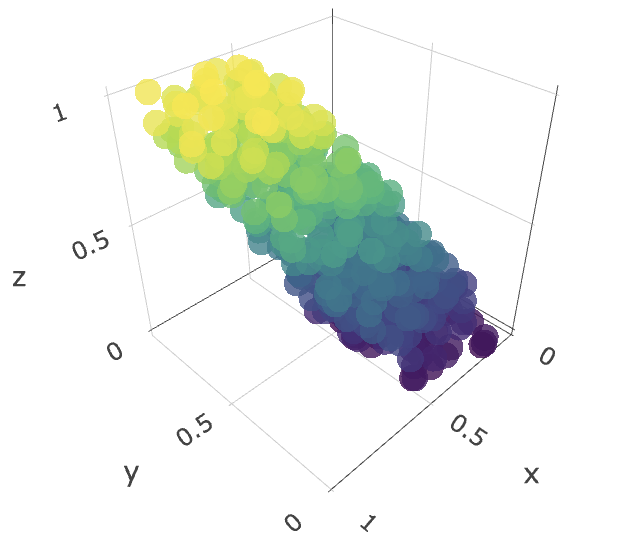} &  \includegraphics[width=11em]{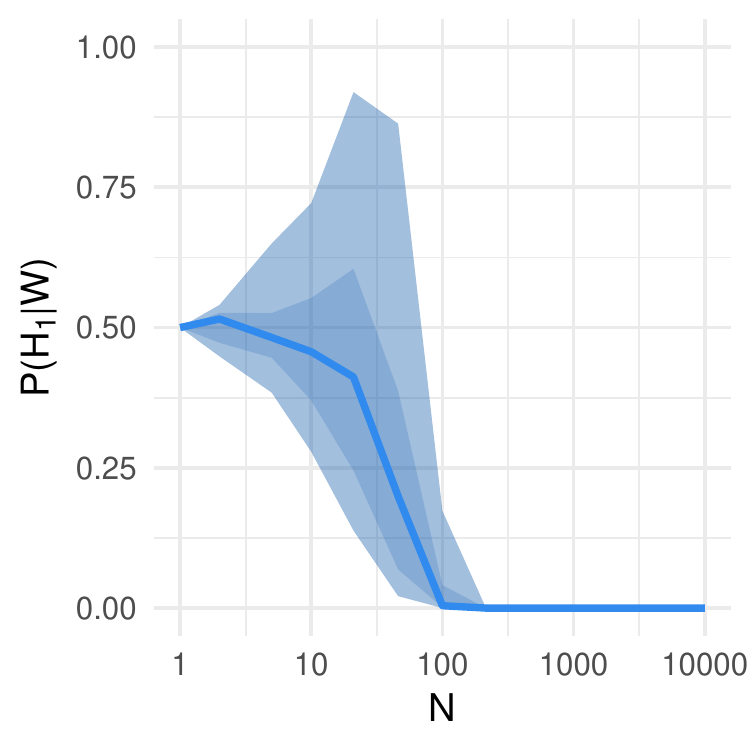} \\
\textbf{3}& \vspace{-2em} \small{
$\begin{array}{c} X,Y \sim \mathcal{U}(0,1) \\
	W \sim \mathcal{N}(0,0.5) \\[3pt] Z = \dfrac{1}{\pi} \text{tan}^{-1}\!\!\left(\dfrac{Y}{X} + W\right) + \dfrac{1}{2}\end{array}$ }
& \includegraphics[width=12em]{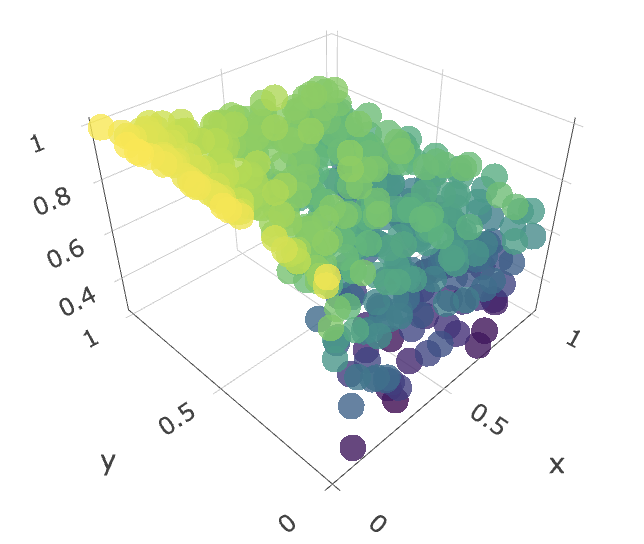}  &
 \includegraphics[width=11em]{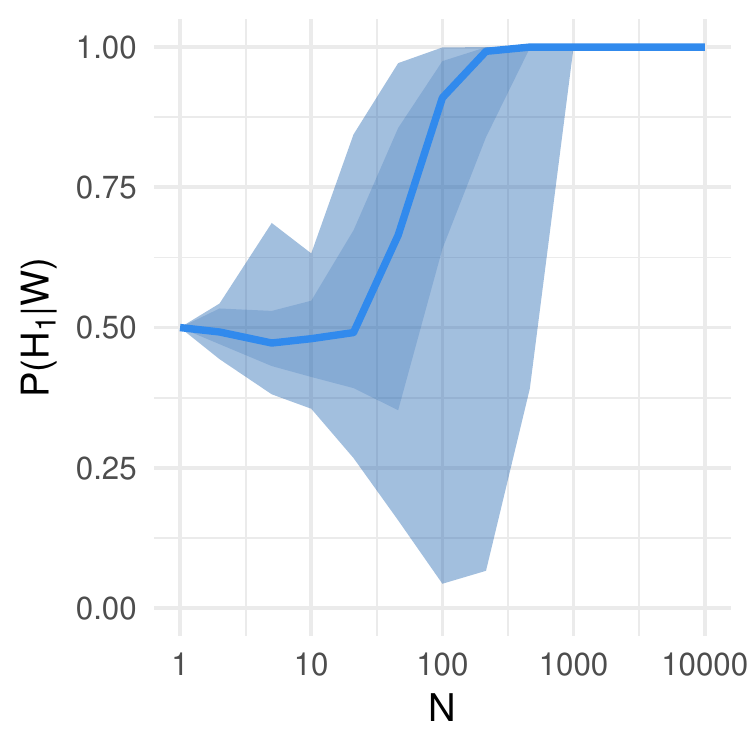}  \\ 
\textbf{4}& \vspace{-5em} \small{
$\begin{array}{c} (X,Y) \sim \mathcal{N}\!\left
	(\left(\begin{smallmatrix}0.5\\0.5\end{smallmatrix}\right)\!,\!\frac{1}{20}\!\left(\begin{smallmatrix}2 & 1 \\ 1 & 2 \end{smallmatrix}\right)\right) \\[-2pt]
	\hspace{1.4cm} \times \mathbf{1}\!\left[(X,Y)\in[0,1]^2\right] \\
	B \sim \mathrm{Bernoulli}(0.9) \\
	\hspace{-1.4cm}Z  \sim \mathcal{N}\!\left
	(0.5,0.5\right)\\[-3pt]
	\hspace{2cm} \times\mathbf{1}\left[Z\in[0,1]\vphantom{^2}
	\right] \\[-3pt]
	\hspace{1.2cm} \times (B+(1-B)XY)
	\end{array}$ }
& \includegraphics[width=12em]{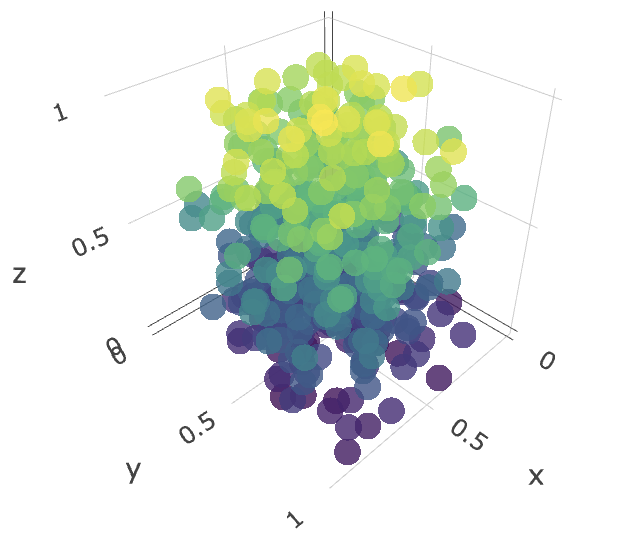} &
 \includegraphics[width=11em]{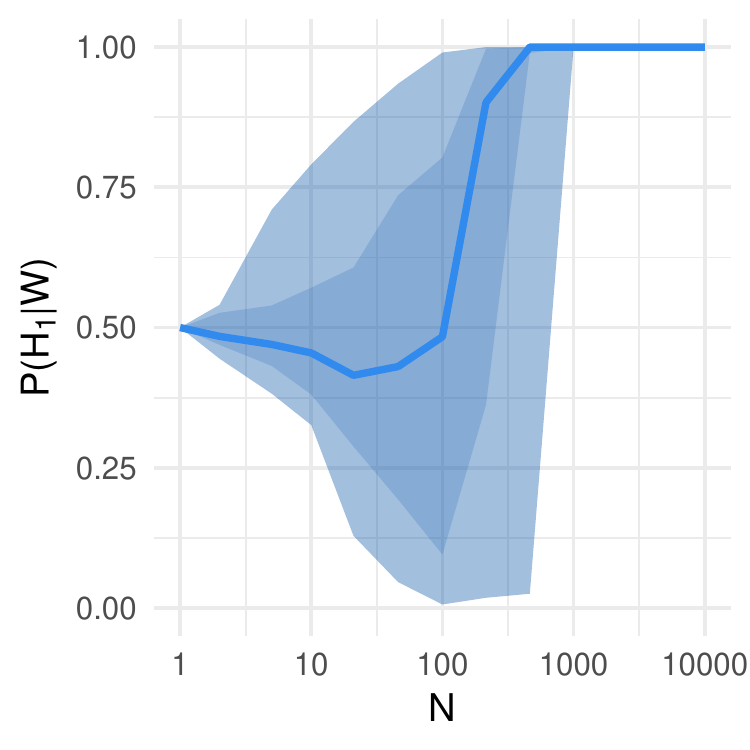} 
 \end{tabular}
}
\captionsetup{width=.94\linewidth}
\caption{Application of the proposed Bayesian testing procedure to four synthetic datasets supported on $[0,1]^3$, chosen such that all combinations of unconditional and conditional dependence/independence are represented. The final column gives the ensemble of probabilities of conditional dependence $p(H_1|W)$ output by the test over 100 repetitions at varying values of data size $N$, with the blue line representing the median, and the dark and light shaded regions representing the (25,75)-percentile and (5,95)-percentile ranges respectively.} \label{table: 4 examples}
\end{figure}
    }
  }{}
}

\title[A Bayesian nonparametric test for conditional independence]{A Bayesian nonparametric test for \\ conditional independence}

\author[Onur Teymur and Sarah Filippi]{}

\email{o@teymur.uk}
\email{s.filippi@imperial.ac.uk}
\thanks{Supported by ESPRC grant EP/R013519/1}

\begin{document}
\maketitle

\centerline{\scshape Onur Teymur and Sarah Filippi}
\medskip
{\footnotesize
\centerline{Department of Mathematics}
\centerline{Imperial College London}
}

\bigskip

\begin{abstract}
This article introduces a Bayesian nonparametric method for quantifying the relative evidence in a dataset in favour of the dependence or independence of two variables conditional on a third. The approach uses P\'{o}lya tree priors on spaces of conditional probability densities, accounting for uncertainty in the form of the underlying distributions in a nonparametric way. The Bayesian perspective provides an inherently symmetric probability measure of conditional dependence or independence, a feature particularly advantageous in causal discovery and not employed in existing procedures of this type.
\end{abstract}

\section{INTRODUCTION}
The random variables $X$ and $Y$ are conditionally independent given $Z$ (written \mbox{$X \indep Y\;|\;Z$}) if and only if the following relation holds between their conditional densities, for all possible realised values $z$ of $Z$:
\begin{equation}
	p_{XY|Z}(x,y|z) =\, p_{X|Z}(x|z)\cdot p_{Y|Z}(y|z)
	\label{eq: conditional densities}
\end{equation}
A common problem in the analysis of multi-variable datasets
is that of assessing whether or not this relation is true for a given triple of variables. Typically, the setting is that the three densities in \eqref{eq: conditional densities}---and the marginal density $p_Z(z)$---are all unknown \emph{a priori}, but we have a finite set of data $W := \{(X_i,Y_i,Z_i)\ ;\ i = 1,\dots,N\}$ assumed to be drawn from the joint measure $p_{XYZ}$ induced by $(X,Y,Z)$. Notably, this type of analysis is a key component in most common approaches to causal discovery \citep{pearl09}. 

Testing for conditional independence with finite data is, however, known to be a hard problem in general. This is particularly true if the unknown densities are assumed continuous and modelled nonparametrically. In such a setting, a test for conditional independence with desirable statistical properties cannot in general be constructed \cite{bergsma04,shah18}. Nonetheless, many tests exist and are commonly used in practice, despite their various theoretical deficiencies. A classic approach is to form a test statistic from the partial correlation coefficient \citep{fisher24}. This vanishes if $X \indep Y\ |\ Z$, but only under the strong assumptions that all variables are Gaussian and all dependences linear. Only limited extensions for non-Gaussian variables \cite{harris13,ramsey14} and for nonlinear dependences \cite{hoyer09,peters14a} exist. Other approaches include combining a series of unconditional independence tests on the response variables $(X,Y)$ conditional on multiple individual values $z$ of $Z$  \citep{margaritis05,huang10}; tests based on measures of statistical distance between estimates of the conditional densities $p_{X|Z}$ and $p_{X|YZ}$, which are zero if and only if $X \indep Y\ |\ Z$ \citep{su07,su08}; tests based on estimation of the conditional mutual information of $X$ and $Y$ given $Z$ \citep{kunihama16, runge17, saad17}; permutation-type tests \citep{candes18,berrett20} that require knowledge of or estimation of $p_{X|Z}$; and a large range of kernel-based methods \citep{fukumizu07,zhang12,doran14,zhang17,strobl18} typically designed with the aim of dealing with high-dimensional or sparse problems more effectively.

All of the methods described in the previous paragraph are frequentist by construction, in that they derive a test statistic and construct a hypothesis test based on either a known null distribution, an asymptotic approximation to it, or by using some other strategy such as a permutation test. In the latter case, the issue is complicated by the fact that permutation tests are not easy to design in the setting of conditional independence testing with continuous $Z$ variable, an issue addressed by a range of modified non-uniform permutation tests \cite{runge17,candes18,berrett20}.

Whichever specific method is used, Peters et al. \cite[\S7.2.1]{peters17} point out one possible problem with relying on a frequentist testing procedure for causal inference, namely that ``all causal discovery methods that are based on conditional independence tests draw conclusions both from dependences and independences''. This reminds us that classical hypothesis testing is inherently asymmetric. Specifically, it is often necessary to detect situations in which the data are `in favour of the null hypothesis' of conditional independence---this is how the PC algorithm \cite{spirtes91} determines which edges to remove in the process of recovering a causal graph. However, doing so subtly abuses the classical hypothesis testing framework, in which one cannot directly compute evidence in favour of the null hypothesis. 

Bayesian hypothesis testing circumvents this issue. To the best of our knowledge there is only a very limited existing literature in Bayesian testing for conditional independence---a method for the case of Gaussian random variables only, for  which conditional independence is equivalent to zero partial correlation \cite{giudici95}.\footnote{There are algorithms among  those surveyed in this section that can be viewed as a `halfway house' towards the Bayesian ideal. In \cite{kunihama16}, for instance, the authors derive a posterior distribution over the conditional mutual information between $X$ and $Y$ given $Z$, which they treat as random in a Bayesian manner. However the output of their method does not directly provide a posterior probability in favour of one of the competing hypotheses.} In this paper, we propose the first Bayesian nonparametric approach for conditional independence testing. The procedure produces a probabilistic measure of the relative evidence in a dataset for dependence or independence of two random variables $X$ and $Y$ conditionally on a third variable $Z$. The nonparametric approach permits the computation of such a probabilistic measure without assuming a known form for the underlying conditional distribution $p_{XY|Z}$. Following Filippi \& Holmes \cite{filippi17}, who construct a Bayesian nonparametric test for (unconditional) independence, we use P\'{o}lya tree priors to model the unknown data-generating distributions.

\subsection{Bayesian nonparametric hypothesis testing}
Recall that we have a dataset $W := \{(X_i,Y_i,Z_i): i=1,\dots,N\}$ and wish to compare two competing hypotheses $H_0$ and $H_1$, with $H_0$ the hypothesis of conditional independence \mbox{and $H_1$ the contrary.}
\begin{equation}
\begin{aligned}
H_0 &: X \indep Y\ |\ Z \\
H_1 &: X \notindep Y\ |\ Z
\end{aligned}
\end{equation}
Our aim is to quantify the relative evidence for these hypotheses in the dataset $W$, which is naturally measured by the posterior probabilities $p(H_0|W)$ and \mbox{$p(H_1|W)$}. 

To evaluate these posterior probabilities,  we use the Bayes Factor \citep{kass95}, defined as the ratio of the marginal likelihoods of two conditional data-generating models.
\begin{equation}
    \mathrm{BF}(H_0,H_1) = \frac{p_{XY|Z}(W|H_0)}{p_{XY|Z}(W|H_1)} \label{eq:BF definition}
\end{equation}
With the prior probabilities of the two hypotheses denoted by $p(H_0)$ and $p(H_1)$, we can use this to derive the posterior probability of $H_1$ as
\begin{equation}
    p(H_1|W) = \frac{1}{1+\mathrm{BF}(H_0,H_1)p(H_0)p(H_1)^{-1}}\label{eq:pH1}
\end{equation}
The ratio of marginal likelihoods on the right-hand side of \eqref{eq:BF definition} can be expanded by factorising the numerator. This is simply an application of the definition of conditional independence given by \eqref{eq: conditional densities}.
\begin{equation}
\frac{p_{XY|Z}(W|H_0)}{p_{XY|Z}(W|H_1)}
 = \frac{p_{X|Z}(W|H_0)p_{Y|Z}(W|H_0)}{p_{XY|Z}(W|H_1)} 
 \label{eq:BF}
\end{equation}
In the remainder we suppress the explicit marking of the models $H_0$ and $H_1$ since the subscripts now make clear which of the three terms belongs to which model.

We now follow a Bayesian nonparametric approach to accommodate the uncertainty in the form of the three unknown conditional densities on the right-hand side of \eqref{eq:BF}. For a domain $\Omega$, we denote by $\mathcal{M}(\Omega)$ the space of all probability measures on $\Omega$. Consider first the two-dimensional conditional density $p_{XY|Z}$ (corresponding to hypothesis $H_1$) with $X$, $Y$ and $Z$ all univariate real  random variables. The Bayesian nonparametric approach entails placing a functional prior $\pi$ on \mbox{$\mathcal{M}(\mathbb{R}^2 \times \mathbb{R})$}---individual elements of which we call $q(\cdot|\cdot)$---incorporating the data $W$ through a likelihood function $\mathcal{L}$, then marginalising over $\mathcal{M}(\mathbb{R}^2 \times \mathbb{R})$ such that the \emph{conditional marginal likelihood} is given by
\begin{equation}\begin{aligned}
	p_{XY|Z}(W) &= \int_{\mathcal{M}(\mathbb{R}^2 \times \mathbb{R})} \mathcal{L}(W;q)\,d\pi(q) \\
	&=\int_{\mathcal{M}(\mathbb{R}^2 \times \mathbb{R})}\:\prod_{i=1}^N q(X_i,Y_i|Z_i)\,d\pi(q)\;;
	\label{eq: marginal conditional likelihood}
\end{aligned}\end{equation}
We refer the reader to the comprehensive textbook treatments in \cite{ghosal17,ghosh03} for further details on the basic principles of the Bayesian nonparametric approach. The same procedure is now applied to the one-dimensional conditional densities $p_{X|Z}$ and $p_{Y|Z}$ with $\pi$, $q \in \mathcal{M}(\mathbb{R} \times \mathbb{R})$ and $\mathcal{L}$ replaced by their one-dimensional analogues. 

Since we wish to assume that the random variables are all continuous, we select $\pi$ to be from the P\'{o}lya tree family of priors. These priors are supported on the entire space of probability measures $\mathcal{M}(\Omega)$ \citep[Thm. 3.3.6]{ghosh03} and can be designed to ensure that individual samples $q$ are absolutely continuous with probability one. Furthermore, they have the advantage that the marginal likelihood in \eqref{eq: marginal conditional likelihood} is tractable, in contrast to other nonparametric models of continuous random variables such as the Dirichlet Process Mixture \citep{escobar95}. 

The specific model we use is a modified version of the conditional Optional P\'{o}lya tree (cond-OPT) of Ma \cite{ma17}, also incorporating ideas from the finite P\'{o}lya tree of Lavine \cite{lavine94} and the multi-dimensional P\'{o}lya tree of Paddock \cite{paddock99}. We review these models in the coming sections. The first constructions we explore are designed for modelling random \emph{unconditional} density functions $q(\cdot)$; later we will see how to build upon these to model random \emph{conditional} density functions $q(\cdot|\cdot)$.

\section{P\'{O}LYA TREES} \label{sec: Polya trees}
\begin{figure}[t]
\begin{center}
\begin{tikzpicture}[scale=1.2]
\draw[line] (0,7.9) -- (0,8.1); 
\draw[line] (8,7.9) -- (8,8.1); 
\draw[line] (0,8) -- (8,8);
\node at (4,8.3) {$\Omega$};
\node at (0,7.7) {$0$};
\node at (8,7.7) {$1$};
\draw[line] (0,6) -- (8,6);
\draw[line] (0,5.9) -- (0,6.1); 
\draw[trans] (4,7.9) -- (2,6.6) node[midway,in front of path,fill=white] {$\theta_0$};
\node at (2,6.3) {$C_0$};
\draw[line] (4,5.9) -- (4,6.1); 
\draw[trans] (4,7.9) -- (6,6.6) node[midway,in front of path,fill=white] {$\theta_1$};
\node at (6,6.3) {$C_1$};
\draw[line] (8,5.9) -- (8,6.1);
\node at (0,5.7) {$0$}; 
\node at (4,5.7) {$0.5$}; 
\node at (8,5.7) {$1$}; 
\draw[line] (0,4) -- (8,4);
\draw[line] (0,3.9) -- (0,4.1); 
\draw[trans] (2,5.9) -- (1,4.6) node[pos=0.5,in front of path,fill=white] {$\theta_{00}$};
\node at (1,4.3) {$C_{00}$};
\draw[line] (2,3.9) -- (2,4.1); 
\draw[trans] (2,5.9) -- (3,4.6) node[pos=0.5,in front of path,fill=white] {$\theta_{01}$};
\node at (3,4.3) {$C_{01}$};
\draw[line] (4,3.9) -- (4,4.1); 
\draw[trans] (6,5.9) -- (5,4.6) node[pos=0.5,in front of path,fill=white] {$\theta_{10}$};
\node at (5,4.3) {$C_{10}$};
\draw[line] (6,3.9) -- (6,4.1);
\draw[trans] (6,5.9) -- (7,4.6) node[pos=0.5,in front of path,fill=white] {$\theta_{11}$};
\node at (7,4.3) {$C_{11}$}; 
\draw[line] (8,3.9) -- (8,4.1);  
\node at (0,3.7) {$0$};  
\node at (8,3.7) {$1$}; 
\draw[line] (0,2) -- (8,2);
\draw[line] (0,1.9) -- (0,2.1); 
\draw[trans] (1,3.9) -- (1/2,2.6) node[midway,in front of path,fill=white] {$\theta_{000}$};
\node at (1/2,2.4) {$C_{000}$};
\draw[line] (1,1.9) -- (1,2.1); 
\draw[trans] (1,3.9) -- (1+1/2,2.6);
\node at (1+1/2,2.4) {$C_{001}$};
\draw[line] (2,1.9) -- (2,2.1); 
\draw[trans] (3,3.9) -- (2+1/2,2.6) node[midway,in front of path,fill=white] {$\;\theta_{010}\;$};
\node at (2+1/2,2.4) {$C_{010}$};
\draw[line] (3,1.9) -- (3,2.1);
\draw[trans] (3,3.9) -- (3+1/2,2.6);
\node at (3+1/2,2.4) {$C_{011}$}; 
\draw[line] (4,1.9) -- (4,2.1); 
\draw[trans] (5,3.9) -- (4+1/2,2.6)
node[midway,in front of path,fill=white] {$\;\theta_{100}\;$};
\node at (4+1/2,2.4) {$C_{100}$};
\draw[line] (5,1.9) -- (5,2.1);
\draw[trans] (5,3.9) -- (5+1/2,2.6);
\node at (5+1/2,2.4) {$C_{101}$};  
\draw[line] (6,1.9) -- (6,2.1); 
\draw[trans] (7,3.9) -- (6+1/2,2.6)
node[midway,in front of path,fill=white] {$\;\theta_{110}\;$};
\node at (6+1/2,2.4) {$C_{110}$};
\draw[line] (7,1.9) -- (7,2.1); 
\draw[trans] (7,3.9) -- (7+1/2,2.6);
\node at (7+1/2,2.4) {$C_{111}$};
\draw[line] (8,1.9) -- (8,2.1);  
\node at (0,1.7) {$0$}; 
\node at (8,1.7) {$1$}; 
\end{tikzpicture}

\end{center}
\captionsetup{width=0.88\linewidth}
\caption{Construction of a P\'{o}lya tree distribution on $\Omega = [0,1]$. From each set $C_\ast$, a particle of probability mass passes to the left with (random) probability $\theta_{\ast0}$ and to the right with probability $\theta_{\ast1} = 1-\theta_{\ast0}$, with all $\theta_\ast$ being independently Beta-distributed as described in the main text.} \label{fig: tree}
\end{figure}
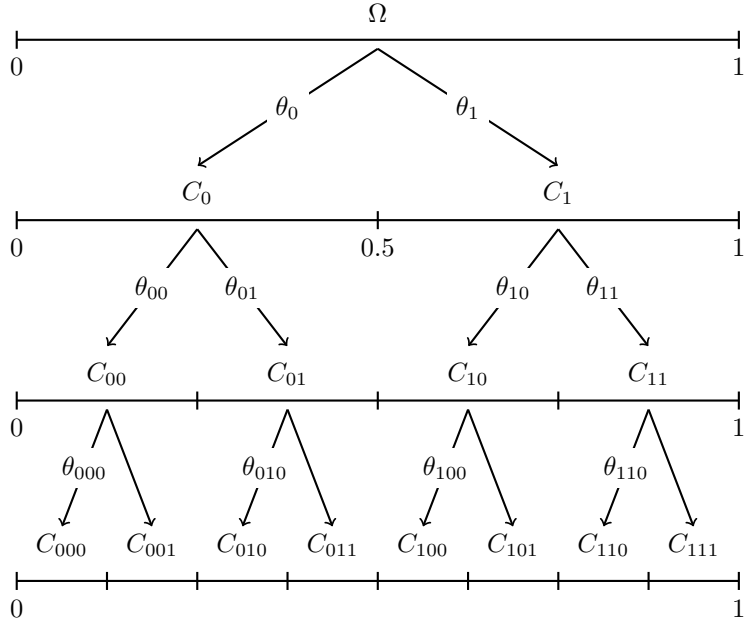

The classical (unconditional) P\'{o}lya tree (PT) \citep{lavine92a,mauldin92,lavine94} essentially defines a random probability measure over a one-dimensional domain \mbox{$\Omega \subseteq \mathbb{R}$}. The most familiar construction proceeds by recursive binary partitioning of $\Omega$ and at each step the assigning of probability mass to the two child sets of a set \mbox{$C \subseteq \Omega$} by means of independent Beta-distributed random branching variables $\theta$. This results in a tree structure, similar to that shown in Figure \ref{fig: tree}. Constructed this way, it is helpful to think of the P\'{o}lya tree as a random histogram on $\Omega$ or, for parameter choices which result in continuous distributions almost surely, a random density function. A particle of probability mass can be thought of as cascading down the tree, with the direction it takes at each binary split \mbox{determined by the random parameters $\theta$.}

More precisely, let $q$ denote a random probability density\footnote{We abuse notation slightly by writing $q$ both for the measure and for its density, the existence of which is always assumed.} on $\Omega$ and $\pi$ a measure over $\mathcal{M}(\Omega)$.
Consider a partitioning of $\Omega$ in two disjoint sets $C_0$ and $C_1$, define the random branching probability $\theta_0 \equiv q(C_0) \sim \mathrm{Be}(\alpha_1,\alpha_1)$ for some $\alpha_1>0$. It follows that $\theta_1 \equiv q(C_1) = 1-\theta_0$. Note that in general, the two parameters of this Beta distribution need not be the same, though this symmetrising simplification is common and we adopt it. Indeed, we take the parameters constant within each level of the tree; the subscript on $\alpha_j$ denotes this level. Continue in this fashion, with $C_0 = C_{00}\cup C_{01}$, $C_{00}\cap C_{01}=\emptyset$ and $\theta_{00} \equiv q(C_{00}|C_0) \sim \mathrm{Be}(\alpha_{2},\alpha_{2})$, $\theta_{000} \equiv q(C_{000}|C_{00}) \sim \mathrm{Be}(\alpha_{3},\alpha_{3})$ and so on recursively, with each independent Beta random variable $\theta_{\ast}$ determining the probability that the particle enters the set $C_{\ast}$ at the next level of the tree. 

We write $\varepsilon_i$ for a (single) element of the set $\{0,1\}$, $\varepsilon^j \equiv \varepsilon_1\varepsilon_2\dots\varepsilon_j$ for a length-$j$ word from the set $\{0,1\}^j$, $\varepsilon^j0$ and $\varepsilon^j1$ for the appending of respectively a single $0$ or $1$ onto the end of $\varepsilon^j$, and $E^j$ for the set of all length-$j$ $\{0,1\}$--words. We further write $\varepsilon^\ast$ for an element of the set $E^\ast \equiv \bigcup_{j=1}^\infty E^j$ of all possible $\{0,1\}$--words of \emph{any} finite length. The measure of a set $C_{\varepsilon_1\varepsilon_2\dots\varepsilon_j}$ can then be written as
\begin{equation}
	q(C_{\varepsilon_1\varepsilon_2\dots\varepsilon_j}) = \prod_{i=1}^j q(C_{\varepsilon_1\varepsilon_2\dots\varepsilon_i} |C_{\varepsilon_1\varepsilon_2\dots\varepsilon_{i-1}})
	\label{eq: set probability}
\end{equation}
Taking the infinite limit of tree depth $j$, it can be shown that the set of finite unions of intervals of the form $C_{\varepsilon^\ast}$ generates the Borel $\sigma$-algebra on $\Omega$. With $q$ constructed in this fashion, the measure $\pi(q)$ is a P\'{o}lya tree.

Under certain conditions on the parameters $\alpha$, the P\'{o}lya tree assigns positive probability to the Kullback--Leibler neighbourhood of any element of the space $\mathcal{M}(\Omega)$. Furthermore, these elements can be made to be absolutely continuous with respect to Lebesgue measure \cite{lavine94}. Specifically, the parameter choice $\alpha_j = cj^2$, with $c>0$ and $j$ the level of the set in question within the tree, satisfies this condition and ensures that samples from the PT are almost surely continuous. We use this choice throughout our simulations and provide a discussion and robustness analysis for the setting of the constant $c$ in Section \ref{sec: hyperparameters}.

The P\'olya tree just defined is supported on a one-dimensional domain, but a multi-dimensional extension---in which sets $C\subseteq\Omega^d$ are binary-divided in each of $d$ dimensions simultaneously at each step---is considered by Paddock  \cite{paddock99}. In this construction, the children of $C $ are assigned probability mass by means of Dirichlet-distributed random variables $\theta$ supported on the $2^d$-dimensional simplex, generalising the Beta-distributed $\theta$ of the one-dimensional PT.\footnote{Recall that if $\theta_0 \sim \text{Beta}(\alpha,\alpha)$ and $\theta_1 = 1-\theta_0$, then $\theta \equiv (\theta_0,\theta_1) \sim \text{Dirichlet}(\alpha,\alpha)$.} Indices $\varepsilon_i$ now take values in the expanded set $\{0,1,\dots,2^d-1\}$, we have $(\theta_{\varepsilon^{j-1}0},\dots,\theta_{\varepsilon^{j-1}(2^d-1)})\sim\mathrm{Dir}(\alpha_j,\dots,\alpha_j)$, and the sets $E^j$ and $E^\ast$ are redefined accordingly. Note that, by definition, $\sum_{k=0}^{2^d-1}\theta_{\varepsilon^{j-1} k} = 1$. 

\subsection{Bayesian inference with P\'olya trees}
P\'{o}lya trees benefit from the conjugacy of the Binomial and Beta (in the multi-dimensional setting: the Multinomial and Dirichlet) distributions, allowing a simple expression to be derived for the posterior measure over $\mathcal{M}(\Omega)$ after data $X_{1:N} \equiv \{X_1,\dots,X_N\}$ have been observed. Let $\{\theta\}$ be the collection of all $\theta_{\varepsilon^\ast}$, and $\Pi_{\Omega^d}$ the set of all $C_{\varepsilon^\ast}$ arising in the recursive partitioning procedure. Then the density of a point $\mathbf{x} \in \Omega^d$ is given by
\begin{equation}
	q(\mathbf{x}|\{\theta\},\Pi_{\Omega^d}) = \prod_{j=1}^\infty  \prod_{\varepsilon^j \in E^j}
	\theta_{\varepsilon^{j}}^{\mathbb{1}[\mathbf{x} \in C_{\varepsilon^{j}}]}
	\label{eq: MPT density function}
\end{equation} 

This equation can be viewed loosely as the limiting case of \eqref{eq: set probability} and unpacked by noting that the conjunction of the product over level-$j$ indices $\varepsilon^j$ and the indicator function in the exponent zeroes all contributions from parameters $\theta_{\varepsilon^j}$ not on the path within the tree that leads to $\mathbf{x}$.

A critical point is that in the classical PT model, exact calculation of quantities such as \eqref{eq: MPT density function} theoretically requires infinite computation, since the tree is of unlimited depth. It is therefore common in practice to truncate the calculation at a finite tree depth $J$. These truncated (also called `finite' or `partially-specified') P\'{o}lya trees (TPT) are discussed by Lavine \cite{lavine94} and Mauldin et al. \cite{mauldin92}. While full Kullback--Leibler support over $\mathcal{M}({\Omega})$ is no longer guaranteed, bounds on the pointwise error of the posterior measure \citep{lavine94} and $L_1$ error of the predictive density \citep{hanson06} are available.  Hanson \& Johnson \cite{hanson02} formalise the definition of the TPT by specifying a base measure $\mu$ that the `leaf' sets at the bottom level $J < \infty$ are taken to follow. If this base measure is uniform then the TPT outputs piecewise-constant measures (ie. random histograms). We follow this approach in the simulations in Section \ref{sec: examples}, but other base measures can be used---for example if $\Omega$ is unbounded and $\mu$ is taken to be a $d$-dimensional Gaussian measure \cite{hanson06}. The density function for the multivariate TPT is given by
\begin{equation}
	q(\mathbf{x}|\{\theta\},\Pi_{\Omega^d},\mu) = \sum_{\varepsilon^J\in E^J}\!\!\!\frac{\mu(\mathbf{x})\mathbb{1}[\mathbf{x}\in C_{\varepsilon^J}]}{\mu(C_{\varepsilon^J\!})} \prod_{j=1}^{J-1} \prod_{\varepsilon^j\in E^j} \theta_{\varepsilon^{j}}^{\mathbb{1}[\mathbf{x} \in C_{\varepsilon^{j}}]}
	\label{eq: TPT density function}
\end{equation} 
The first fraction in this equation is the normalised base density of the point $\mathbf{x}$ within its level-$J$ set. 

We now combine the prior with the likelihood. Conjugacy not only means that the posterior is itself a P\'{o}lya tree, but also that the branching variables $\{\theta\}$ can easily be marginalised. Assuming henceforth that $\mu$ is indeed uniform, this gives the TPT marginal likelihood
\begin{equation}
\begin{aligned}
	p_X(\mathbf{X}_{1:N}|\{\alpha\},\Pi_{\Omega^d},\mu) &= \int \prod_{i=1}^N q(\mathbf{X}_i|\{\theta\},\Pi_{\Omega^d},\mu) p (\{\theta\}|\{\alpha\})\ d\{\theta\} \\ &= \frac{1}
	{2^{dJn}}\prod_{j=1}^{J-1} \frac{\Gamma(2^d\alpha_j) \cdot \prod_{\varepsilon^j \in E^j}\Gamma(\alpha_j + n_{\varepsilon_j}(\mathbf{X}_{1:N}))} {\Gamma(\alpha_j)^{2^d} \cdot \Gamma({2^d}\alpha_j + \sum_{\varepsilon^j \in E^j} n_{\varepsilon_j}(\mathbf{X}_{1:N}) )}
	\label{eq: first MPT marg lik}
\end{aligned}
\end{equation}
Here, $n_{\varepsilon_j}(\mathbf{X}_{1:N})$ counts the number of data $\mathbf{X}_{1:N}$ in the set $\varepsilon_j$. It is then possible to derive the predictive distribution, and using this, an alternative expression for the marginal likelihood that is easier to work with in practice.
\begin{equation}
	p_X(\mathbf{x}|\mathbf{X}_{1:N},\{\alpha\},\Pi_{\Omega^d},\mu) = \prod_{j=1}^J\frac{2^d\alpha_j+2^d n_j(\mathbf{x};\mathbf{X}_{1:N})}{2^d\alpha_j+n_{j-1}(\mathbf{x};\mathbf{X}_{1:N}))}
	\label{eq: MPT posterior density}
\end{equation}
\begin{equation}
	p_X(\mathbf{X}_{1:N}|\{\alpha\},\Pi_{\Omega^d},\mu) = \prod_{i=2}^N \prod_{j=1}^J\frac{2^d\alpha_j+2^d n_j(\mathbf{X}_i;\mathbf{X}_{1:i-1})}{2^d\alpha_j+n_{j-1}(\mathbf{X}_i;\mathbf{X}_{1:i-1})}
	\label{eq: MPT marginal likelihood telescope}
\end{equation}
In these equations, $n_{j}(\mathbf{x};\mathbf{X}_{1:N})$ counts the number of data in $\{\mathbf{X}_1,\dots,\mathbf{X}_{N}\}$ that are at the same level-$j$ set as $\mathbf{x}$, ie. $n_{j}(\mathbf{X}_i;\mathbf{X}_{1:i-1}) = \sum_{k=1}^{i-1}\mathbb{1}[\mathbf{X}_k\in C_{\varepsilon^j}]\mathbb{1}[\mathbf{X}_i\in C_{\varepsilon^j}]$.
 
\subsection{P\'{o}lya tree models for conditional distributions} \label{sec: OPT}

In this section we describe how the canonical P\'olya tree construction described in Section \ref{sec: Polya trees} can be extended to model \emph{conditional} distributions. Doing so first requires a notion of randomised partitioning called `optional stopping'.
This was first proposed by Wong \& Ma \cite{wong10} as an alternative solution to the problem of ensuring that computation time in PT modelling be made almost surely finite. In this paradigm, called the Optional P\'{o}lya tree (OPT), the partitioning of $\Omega$ is augmented at each step by the drawing of independent Bernoulli-distributed stopping variables $S$. For a set $C_\ast$ arising in the partitioning of $\Omega$, if the corresponding $S_\ast$ is equal to $1$ then $C_\ast$ is divided no further and a uniform distribution is placed on it. If $S_\ast = 0$ then a binary split takes place as usual. This outcome of this procedure is a (random) partition of varying granularity across the domain.

As long as the Bernoulli parameter $\rho$ controlling the probability $\mathrm{Pr}(S_\ast = 1)$ is uniformly greater than 0 for all sets $C_\ast$, it is easy to see that this algorithm will result in all of $\Omega$ (but for a set of measure zero) being `stopped' in finite time with probability one. The additional randomness introduced by this partitioning procedure is itself marginalised to give quantities analogous to \eqref{eq: MPT posterior density} and \eqref{eq: MPT marginal likelihood telescope} that can be calculated in finite time. Given certain further technical conditions, a full-support result akin to that for the classical PT is also available.

The optional stopping principle is then further leveraged in Ma \cite{ma13,ma17}, in which multi-scale mixtures of OPTs are used as models for conditional probability distributions; this is called the conditional Optional P\'{o}lya tree (cond-OPT) \cite{ma17}. The basic idea is to construct a random conditional density $q(x|z) \in \mathcal{M}(\mathbb{R}\times\mathbb{R})$ by partitioning the predictor space $\Omega_Z$ using the optional-stopping algorithm described above, then for each set $A$ arising from this procedure to construct an independent (unconditional) OPT random density on the response space $\Omega_X$ but using \emph{only} those data $X_i$ whose corresponding $Z_i$ value lies in $A$. Finally, the multiple independent models over $\Omega_X$ are combined in a weighted sum (with the weights determined by the partition of $\Omega_Z$), giving a random conditional density $q(x|z)$.

The measure constructed this way has full (total variation) support on the space $\mathcal{M}(\mathbb{R}\times\mathbb{R})$ of conditional density functions supported on $\Omega_X \times \Omega_Z$ \cite{ma17}, and as such is a direct generalisation of the unconditional P\'{o}lya tree family of models so far discussed, immediately inheriting many of their strengths. This construction for modelling random conditional density functions forms a central part of our work and we describe it in much greater detail in the next section.

\section{A BAYESIAN CONDITIONAL INDEPENDENCE TEST}

Recall that we seek to compare the hypotheses $
	H_0: X\indep Y\;|\;Z$ versus $H_1: X\notindep Y\;|\;Z $.
	
Call the support of $X$, $Y$ and $Z$ respectively $\Omega_X$, $\Omega_Y$ and $\Omega_Z$ and assume that $\Omega := \Omega_X \times \Omega_Y \times \Omega_Z$ is a compact subset of $\mathbb{R}^3$. We will define three nonparametric priors, one for each of the three conditional density functions appearing in \eqref{eq:BF}. Then, by incorporating the data $W$ and marginalising the randomness in the posterior, we will derive the three conditional marginal likelihoods required to calculate \mbox{the Bayes Factor.}

We use $p_{X|Z}$ as our running example; the models for $p_{Y|Z}$ and $p_{XY|Z}$ are the same, with the obvious modifications. The approach consists in first constructing a random partition of $\Omega_Z$ and then, for each partition block $A$, generating the distribution of $X$ conditionally on $Z\in A$ using a truncated P{\'o}lya tree (TPT). The first step is to partition $\Omega_Z$ using the optional-stopping binary recursive partitioning procedure described in Section \ref{sec: OPT}. This produces a \emph{random} partition of $\Omega_Z$---this is an intrinsic feature of this scheme. This additional randomness will itself be marginalised in order to calculate the conditional marginal likelihood $p_{X|Z}(W)$. Following Ma \cite{ma17}, this is done in practice by constructing a \emph{non-random} binary partition $\Pi_{\Omega_Z}$ and performing a recursive calculation on the resulting tree. We now explain this calculation in detail.

For any $A \in \Pi_{\Omega_Z}$, let $W_A = \{(X_i,Y_i,Z_i)\ ;\ i=1,\dots,N : Z_i \in A\}$ be the subset of the data $W$ whose $Z$ component is in $A$, and let $N_A = \left| W_A \right|$ be the cardinality of this set. We also write $X_A$ for the set of $X$ components of $W_A$, and similarly for $Y_A$ and $Z_A$. For each set $A$, we consider a `local' conditional distribution of $X$ given $Z=z$ (which is assumed to be constant across all $z\in A$) and use a TPT prior for this distribution. The `local'
likelihood of the data $X_A$ is therefore given by $$q^0_X(A):=\prod_{i=1}^{N_A}q((X_A)_i|\{\theta\},\Pi_{X,A})\;,$$ where the contributions from individual data points are given by \eqref{eq: TPT density function}, and $\Pi_{X,A}$ denotes the partition that `separates' $X_A$. More precisely, the partition $\Pi_{X,A} \subseteq \Pi_{\Omega_X} (\equiv \Pi_{X,\Omega_Z})$ of $\Omega_X$ is defined such that all leaf sets contain either $0$ or $1$ data point from $X_A$.\footnote{In practice such partition is calculated most efficiently by constructing the most extensive tree $\Pi_{X,\Omega_Z}$ once, then pruning it to find the $\Pi_{X,A}$ for each $A$.}
The full multi-scale conditional likelihood $q_X(A)$ is then determined recursively by drawing stopping variables $S_X(A)$, and calculating $q^0_X(A)$ for all sets $A$ arising in the resulting random partition of $\Omega_Z$. For any set $A_\ast$ which remains unstopped, we call its two children $A_{\ast 0}$ and $A_{\ast 1}$. Then $q_X(A_\ast)$ is given by
\begin{equation}
    q_X(A_\ast) := \begin{cases} q_{X}^0(A_\ast) \quad \text{if $S_X(A_\ast) = 1$,} \\
    q_{X}(A_{\ast0})q_{X}(A_{\ast1}) \quad  \text{if $S_X(A_\ast) = 0$.} \end{cases} \label{eq: q cases}
\end{equation}
Equivalently, this can be written as an additive mixture.
\begin{equation}
    q_X(A_\ast) = S_X(A_\ast)q_X^0(A_\ast) + (1-S_X(A_\ast))q_X(A_{\ast0})q_X(A_{\ast1})
    \label{eq: mixture lik}
\end{equation}
To calculate the conditional \emph{marginal} likelihood, this expression needs to be integrated to marginalise the randomness from both the local likelihoods $\{q^0_X\}$, and the partitioning procedure, determined by $\{S_X\}$. We write the local marginal likelihoods as $\Phi_X^0(A) := p_X(X_A|\{\alpha\},\Pi_{X,A})$, and from equation \eqref{eq: MPT marginal likelihood telescope} we have
\begin{equation}
	\Phi_X^0(A) = \prod_{i=2}^{N_A} \prod_{j=1}^{J_X}\frac{2\alpha_j+2 n_j((X_A)_i;(X_A)_{1:i-1})}{2\alpha_j+n_{j-1}((X_A)_i;(X_A)_{1:i-1})}
\end{equation}
where $J_X$ is the maximum depth of the partition $\Pi_X$. The complete conditional marginal likelihood 
$\Phi_{X}(A) := p_{X|Z}(W_A)$ is then obtained by marginalising the partitioning randomness from \eqref{eq: mixture lik}. Letting $\rho(A_\ast)=\mathrm{Pr}(S_X(A_\ast)=1)$, we have
\begin{equation}
        \Phi_X(A_\ast) = \rho(A_\ast)\Phi_X^0(A_\ast) + (1-\rho(A_\ast))\Phi(A_{\ast0})\Phi(A_{\ast1})
    \label{eq: mixture marg lik}
\end{equation}

This recursion is performed in practice by starting from the leaf sets of the most extensive non-random separating partition $\Pi_{\Omega_Z}$ and applying the following algorithm, until the root $\Omega_Z$ is reached.
\begin{equation}
    \Phi_{X}(A_\ast) := \begin{cases} \Phi_{X}^0(A_\ast) \quad \text{if $A_\ast$ is a leaf set,} \\
    \rho(A_\ast)\,\Phi_{X}^0(A_\ast) + (1 - \rho(A_\ast))\Phi_{X}(A_{\ast 0})\Phi_{X}(A_{\ast 1}) \ \  \text{if not.} \end{cases} \label{eq: phi}
\end{equation}
The value of this function at the root $\Omega_Z$ is the conditional marginal likelihood we require, ie.
\begin{equation}
	p_{X|Z}(W) = \Phi_X(\Omega_Z)\;.
\end{equation}
The variables $\rho(A) \in (0,1)$ function as mixing parameters and we take them to be constant and equal to 0.5 for all sets $A$---we discuss this choice in Section \ref{sec: hyperparameters}. Equation \eqref{eq: mixture marg lik} makes clear the way in which the conditional marginal likelihood $\Phi_X(\cdot)$ is formed of a multi-scale additive mixture of TPT marginal likelihoods $\Phi_X^0(\cdot)$.

The equivalent calculation is undertaken to find $p_{Y|Z}(W) \equiv \Phi_Y(\Omega_Z)$ and---now using the bivariate version of the TPT---$p_{XY|Z}(W) \equiv \Phi_{XY}(\Omega_Z)$. The Bayes Factor \eqref{eq:BF} is then given by
\begin{equation}
\mathrm{BF}(H_0,H_1) = \frac{\Phi_{X}(\Omega_Z)\Phi_{Y}(\Omega_Z)}{\Phi_{XY}(\Omega_Z)} \label{eq: second bayes factor}
\end{equation}
and, the posterior probability of conditional dependence $ p(H_1|W)$ can then be obtained using \eqref{eq:pH1}. The algorithm described in this section is summarised in the pseudocode in Figure \ref{pseudocode}.

\begin{figure}[t]
\fbox{
\parbox{0.91\columnwidth}{
\vspace{0.1cm} \small
\textbf{Bayesian nonparametric test to assess $H_0: X\indep Y\;|\;Z $ vs. $ H_1: X\notindep Y\;|\;Z $}\\\vspace{-0.7em} \hrule \vspace{0.5em}

\textbf{inputs:} data $W = \{(X_i,Y_i,Z_i): i=1,\dots,N\}$; parameters $\rho,c$; \\ \hphantom{\textbf{inputs: }}finite `separating' partitions $\Pi_{\Omega_Z}, \Pi_{\Omega_X}, \Pi_{\Omega_Y} $ and $\Pi_{\Omega_{XY}}$ \\[-0.6em] 

\textbf{for all} $A$ in $\Pi_{\Omega_Z}$ \\
\hspace*{1em} //\textit{partition pruning}\\
\hspace*{1em} $W_A = \{(X_i,Y_i,Z_i) : Z_i \in A\}$ ($W_A \equiv (X_A,Y_A,Z_A)$) \\
\hspace*{1em} construct $\Pi_{X,A}$, $\Pi_{Y,A}$ and $\Pi_{XY,A}$ by pruning $\Pi_{\Omega_X}, \Pi_{\Omega_Y} $ and $\Pi_{\Omega_{XY}}$, \\ \hspace*{2.5em} keeping only those blocks containing $\geq$ 2 data points from $W_A$ \\[-0.6em] 

\hspace*{1em}// \textit{calculate TPT marginal likelihoods } \hfill  \eqref{eq: MPT marginal likelihood telescope}\\
  \hspace*{1em} $\Phi_X^0(A) \leftarrow p_X(X_A|\{\alpha\},\Pi_{X,A})$ \\
  \hspace*{1em} $\Phi_Y^0(A) \leftarrow p_Y(Y_A|\{\alpha\},\Pi_{Y,A})$ \\
  \hspace*{1em} $\Phi_{XY}^0(A) \leftarrow p_{XY}((X_A,Y_A)|\{\alpha\},\Pi_{XY,A})$ \\[-0.6em]
 
// \textit{calculate conditional marginal likelihoods } \hfill  \eqref{eq: phi}\\
\textbf{for all} leaf sets $A$ in $\Pi_{\Omega_Z}$ \\
  \hspace*{1em} $\Phi_X(A) \leftarrow \Phi^0_X(A)$\\
  \hspace*{1em} $\Phi_Y(A) \leftarrow \Phi^0_X(A)$ \\
  \hspace*{1em} $\Phi_{XY}(A) \leftarrow \Phi^0_{XY}(A)$ \\[-0.6em]
  
\textbf{for all} non-leaf sets $A$ in $\Pi_{\Omega_Z}$ with children $A_{0}$ and $A_{1}$ \hfill \eqref{eq: phi} \\
 \hspace*{1em} // \textit{traversal order from leaf sets towards root}\\
  \hspace*{1em} $\Phi_X(A) \leftarrow \rho\Phi^0_X(A) + (1-\rho)\Phi_X(A_{0})\Phi_X(A_{1})$ \\
  \hspace*{1em} $\Phi_Y(A) \leftarrow \rho\Phi^0_Y(A) + (1-\rho)\Phi_Y(A_{0})\Phi_Y(A_{1})$ \\
  \hspace*{1em} $\Phi_{XY}(A) \leftarrow \rho\Phi^0_{XY}(A) + (1-\rho)\Phi_{XY}(A_{0})\Phi_{XY}(A_{1})$ \\[-0.6em]

  \textbf{output:} $\mathrm{BF} \leftarrow \Phi_X(\Omega_Z)\Phi_Y(\Omega_Z)(\Phi_{XY}(\Omega_Z))^{-1}$ \hfill \eqref{eq: second bayes factor}
  
  \vspace{+0.1cm}
}} 
\captionsetup{width=0.60\linewidth}
\caption{ Pseudocode for the proposed Bayesian nonparametric test for conditional independence} \label{pseudocode}
\end{figure}

\section{EXPERIMENTS} \label{sec: examples}

In this section we describe some example experiments to elucidate the operation and output of the proposed approach. We stress once again that the output of our algorithm is a Bayesian posterior probability value $p(H_1|W)$ which is directly interpretable as a ``probability of conditional dependence'', in contrast to previous approaches, which derive or approximate a threshold value for a classical test statistic. This fundamental difference makes direct comparison with existing methods challenging.

\subsection{Synthetic data}

Our first set of experiments uses synthetic datasets constructed by the formulae in the first column of Figure \ref{table: 4 examples}. The measures from which the data are sampled are designed in such a way that every combination of unconditional independence/dependence and conditional independence/dependence is represented. Specifically, in model 1 it holds that $X \indep Y$ as well as $X \indep Y\,|\,Z$; in model 2 we have $X \notindep Y$ but $X \indep Y\,|\,Z$; in model 3 it holds that $X \indep Y$ though $X \notindep Y\,|\,Z$, and in model 4 we have $X \notindep Y$ and $X \notindep Y\,|\,Z$. In each case, $(X,Y,Z)$ are by construction supported on $\Omega_X=\Omega_Y=\Omega_Z=[0,1]$. Example 3-dimensional scatter plots are given for each model in the middle column.

We highlight specifically model 4, for which $X \notindep Y$ and $X \notindep Y\,|\,Z$, though the generating process is a mixture and for 90\% of the data it holds that $X \indep Y\,|\,Z$. Noting definition \eqref{eq: conditional densities} (``for all $z$''), we would like a partial conditional dependence of this type to be detected by a hypothesis test, even if it derives from only a small subset of the data.

We vary the number of data $N$ between $1$ and $10^5$, and for each of several values of $N$ in this range we run 100 repetitions of our procedure using datasets generated by different random seeds. We consider binary recursive partitions of $\Omega_X=\Omega_Y=\Omega_Z=[0,1]$ which at level $j$ have the form 
\begin{equation}
    [0,1]=\bigcup_{k=0}^{2^j-1}\left[\frac{k}{2^j},\frac{k+1}{2^j}\right).
    \label{binary partition def}
\end{equation}
The maximum tree depths $J_Z$ (in the predictor space) and $J_X$, $J_Y$ and $J_{XY}$ (in the response spaces) are all set at $\lceil\log_2(N)\rceil$, following a widely-used rule of thumb \cite{hanson02}. In addition, we assume an equal prior value for both hypotheses, so that  $p(H_0)=p(H_1)=0.5$.

We plot the range of test outputs $p(H_1|W)$ in the right-hand column of Figure \ref{table: 4 examples}, with the blue line representing the median, and the dark- and light-blue shaded regions representing the (25,75)-percentile range and the (5,95)-percentile range respectively. In the low-data limit, the test output $p(H_1|W)$ converges to $0.5$ as expected, indicating reversion to the prior probability $p(H_1)$, while for values of $N$ of $10^4$ and greater the test consistently returns a probability value very close to $0$ or $1$, correctly determining in each case the hypothesis that reflects the ground truth.

In the approximate range $N=10^1$ to $10^3$, a relatively large uncertainty is present in the output. In the case of the two examples for which $X\notindep Y\;|\;Z$ (models 3 and 4), there is a noticeable tendency in this range to falsely favour $H_0$, before $p(H_1|W)$ converges to $1$ correctly as $N>10^4$. This is a manifestation of the natural Occam Factor present in the test, favouring the simpler model $H_0$ where insufficient data exists to conclusively support $H_1$ \cite[\S 28]{mackay03}. The same phenomenon was observed in the unconditional independence testing procedure upon which this work builds \cite{filippi17}.

\begin{figure}[p!]
\vspace{1em}
     \includegraphics[width=\textwidth]{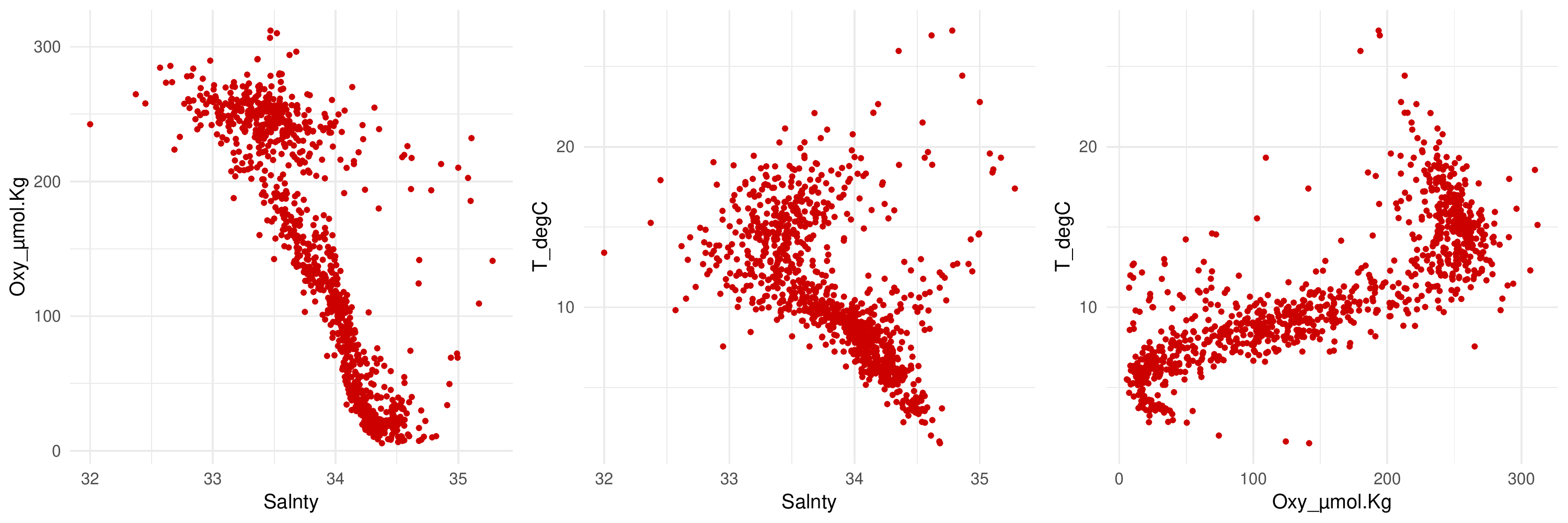}
      \captionsetup{width=.92\linewidth}
      \caption{Marginal scatter plots from the CalCOFI Bottle dataset showing the pairwise relationships between \texttt{Salnty},  \texttt{Oxy\_µmol.Kg} and \texttt{T\_degC}. The nonlinear nature of the dependences is immediately apparent.} \label{figure: calcofi}

\vspace{5em}

\centering
\begin{minipage}{0.42\textwidth}

\begin{tikzpicture}
    \node[shape=circle,draw=black,minimum size=0.3cm,align = center,label={[xshift=0cm, yshift=-0.1cm]\texttt{\small Salnty}}] (A) at (0,1.2*1) {};
    \node[shape=circle,draw=black,minimum size=0.3cm,align = center,label={[xshift=-0.7cm, yshift=-0.2cm]\texttt{\small STheta}}] (B) at (1.2*-0.9510565,1.2*0.309017) {};
    \node[shape=circle,draw=black,minimum size=0.3cm,align = center,label={[xshift=0.7cm, yshift=-0.2cm]\texttt{\small R\_PRES}}] (E) at (1.2*0.9510565,1.2*0.309017) {};
    \node[shape=circle,draw=black,minimum size=0.3cm,align = center,label={[xshift=-0.3cm, yshift=-0.8cm]\texttt{\small Oxy\_µmol}}] (C) at (1.2*-0.5877853,1.2*-0.809017) {};
    \node[shape=circle,draw=black,,minimum size=0.3cm,align = center,label={[xshift=0.3cm, yshift=-0.8cm]\texttt{\small R\_DYNHT}}] (D) at (1.2*0.5877853,1.2*-0.809017) {};

    \path [-,line width=0.5mm] (A) edge node[midway,fill=white,inner sep=2pt] {\footnotesize 0.99}(B);
    \path [-,line width=0.5mm](B) edge node[midway,fill=white,inner sep=2pt]  {\footnotesize 0.99}(C);
    \path [-,line width=0.5mm](A) edge node[midway,fill=white,inner sep=2pt] {\footnotesize 1}(C);
    \path [-,line width=0.5mm](D) edge node[midway,fill=white,inner sep=2pt] {\footnotesize 1} (E); 
    \path [-,line width=0.5mm](B) edge node[pos=0.8,fill=white,inner sep=2pt] {\footnotesize 0.98} (D); 
    \path [-,line width=0.25mm](C) edge node[midway,fill=white,inner sep=2pt] {\footnotesize 0.53} (D); 
    \path [-,line width=0.15mm](B) edge node[pos=0.65,fill=white,inner sep=2pt] {\footnotesize 0.33} (E);
    \path [-,line width=0.05mm](C) edge node[pos=0.75,fill=white,inner sep=2pt] {\footnotesize 0.01} (E);
    \path [-,line width=0.05mm](A) edge node[pos=0.2,xshift=0.3cm,yshift=0.1cm,inner sep=2pt] {\footnotesize 0.04} (D);  
    \node[draw] at (0,-2) {$N=50$};
\end{tikzpicture}
\end{minipage}
\begin{minipage}{0.42\textwidth}
%\centering
\begin{tikzpicture}
    \node[shape=circle,draw=black,minimum size=0.3cm,align = center,label={[xshift=0cm, yshift=-0.1cm]\texttt{\small Salnty}}] (A) at (0,1.2*1) {};
    \node[shape=circle,draw=black,minimum size=0.3cm,align = center,label={[xshift=-0.7cm, yshift=-0.2cm]\texttt{\small STheta}}] (B) at (1.2*-0.9510565,1.2*0.309017) {};
    \node[shape=circle,draw=black,minimum size=0.3cm,align = center,label={[xshift=0.7cm, yshift=-0.2cm]\texttt{\small R\_PRES}}] (E) at (1.2*0.9510565,1.2*0.309017) {};
    \node[shape=circle,draw=black,minimum size=0.3cm,align = center,label={[xshift=-0.3cm, yshift=-0.8cm]\texttt{\small Oxy\_µmol}}] (C) at (1.2*-0.5877853,1.2*-0.809017) {};
    \node[shape=circle,draw=black,,minimum size=0.3cm,align = center,label={[xshift=0.3cm, yshift=-0.8cm]\texttt{\small R\_DYNHT}}] (D) at (1.2*0.5877853,1.2*-0.809017) {};

    \path [-,line width=0.5mm] (A) edge node[midway,fill=white,inner sep=2pt] {\footnotesize 1}(B);
    \path [-,line width=0.5mm](B) edge node[midway,fill=white,inner sep=2pt]  {\footnotesize 1}(C);
    \path [-,line width=0.5mm](A) edge node[midway,fill=white,inner sep=2pt] {\footnotesize 1}(C);
    \path [-,line width=0.5mm](D) edge node[midway,fill=white,inner sep=2pt] {\footnotesize 1} (E); 
    \path [-,line width=0.15mm](B) edge node[pos=0.7,fill=white,inner sep=2pt] {\footnotesize 0.37} (D); 
    \path [-,line width=0.35mm](C) edge node[midway,fill=white,inner sep=2pt] {\footnotesize 0.81} (D); 
    \node[draw] at (0,-2) {$N=100$};
\end{tikzpicture}
\end{minipage}

\vspace{0.7cm}
\begin{minipage}{0.42\textwidth}

\begin{tikzpicture}
    \node[shape=circle,draw=black,minimum size=0.3cm,align = center,label={[xshift=0cm, yshift=-0.1cm]\texttt{\small Salnty}}] (A) at (0,1.2*1) {};
    \node[shape=circle,draw=black,minimum size=0.3cm,align = center,label={[xshift=-0.7cm, yshift=-0.2cm]\texttt{\small STheta}}] (B) at (1.2*-0.9510565,1.2*0.309017) {};
    \node[shape=circle,draw=black,minimum size=0.3cm,align = center,label={[xshift=0.7cm, yshift=-0.2cm]\texttt{\small R\_PRES}}] (E) at (1.2*0.9510565,1.2*0.309017) {};
    \node[shape=circle,draw=black,minimum size=0.3cm,align = center,label={[xshift=-0.3cm, yshift=-0.8cm]\texttt{\small Oxy\_µmol}}] (C) at (1.2*-0.5877853,1.2*-0.809017) {};
    \node[shape=circle,draw=black,,minimum size=0.3cm,align = center,label={[xshift=0.3cm, yshift=-0.8cm]\texttt{\small R\_DYNHT}}] (D) at (1.2*0.5877853,1.2*-0.809017) {};

    \path [-,line width=0.5mm] (A) edge node[midway,fill=white,inner sep=2pt] {\footnotesize 1}(B);
    \path [-,line width=0.5mm](B) edge node[midway,fill=white,inner sep=2pt]  {\footnotesize 1}(C);
    \path [-,line width=0.5mm](A) edge node[midway,fill=white,inner sep=2pt] {\footnotesize 1}(C);
    \path [-,line width=0.5mm](D) edge node[midway,fill=white,inner sep=2pt] {\footnotesize 1} (E); 
    \path [-,line width=0.1mm](B) edge node[pos=0.7,fill=white,inner sep=2pt] {\footnotesize 0.06} (D);
    \node[draw] at (0,-2) {$N=200$};
\end{tikzpicture}
\end{minipage}
\begin{minipage}{0.42\textwidth}

\begin{tikzpicture}
    \node[shape=circle,draw=black,minimum size=0.3cm,align = center,label={[xshift=0cm, yshift=-0.1cm]\texttt{\small Salnty}}] (A) at (0,1.2*1) {};
    \node[shape=circle,draw=black,minimum size=0.3cm,align = center,label={[xshift=-0.7cm, yshift=-0.2cm]\texttt{\small STheta}}] (B) at (1.2*-0.9510565,1.2*0.309017) {};
    \node[shape=circle,draw=black,minimum size=0.3cm,align = center,label={[xshift=0.7cm, yshift=-0.2cm]\texttt{\small R\_PRES}}] (E) at (1.2*0.9510565,1.2*0.309017) {};
    \node[shape=circle,draw=black,minimum size=0.3cm,align = center,label={[xshift=-0.3cm, yshift=-0.8cm]\texttt{\small Oxy\_µmol}}] (C) at (1.2*-0.5877853,1.2*-0.809017) {};
    \node[shape=circle,draw=black,,minimum size=0.3cm,align = center,label={[xshift=0.3cm, yshift=-0.8cm]\texttt{\small R\_DYNHT}}] (D) at (1.2*0.5877853,1.2*-0.809017) {};

    \path [-,line width=0.5mm] (A) edge node[midway,fill=white,inner sep=2pt] {\footnotesize 1}(B);
    \path [-,line width=0.5mm](B) edge node[midway,fill=white,inner sep=2pt]  {\footnotesize 1}(C);
    \path [-,line width=0.5mm](A) edge node[midway,fill=white,inner sep=2pt] {\footnotesize 1}(C);
    \path [-,line width=0.5mm](D) edge node[midway,fill=white,inner sep=2pt] {\footnotesize 1} (E); 
    \node[draw] at (0,-2) {$N=500$};
\end{tikzpicture}
\end{minipage}
\captionsetup{width=0.94\linewidth}
\caption{Example pairwise dependence graphs output by the Bayesian conditional independence test for five variables from the CalCOFI dataset, conditional on \texttt{T\_degC}, for four different sizes of subsample drawn from the complete dataset. The numbers associated with each edge are the posterior probabilities of conditional dependence $p(H_1|W^{(N)})$ and are given to two decimal places; where no edge is shown, this indicates $p(H_1|W^{(N)})<0.005$.} \label{figure: graph}
\end{figure}

\begin{figure}[t]
     \includegraphics[width=0.45\textwidth]{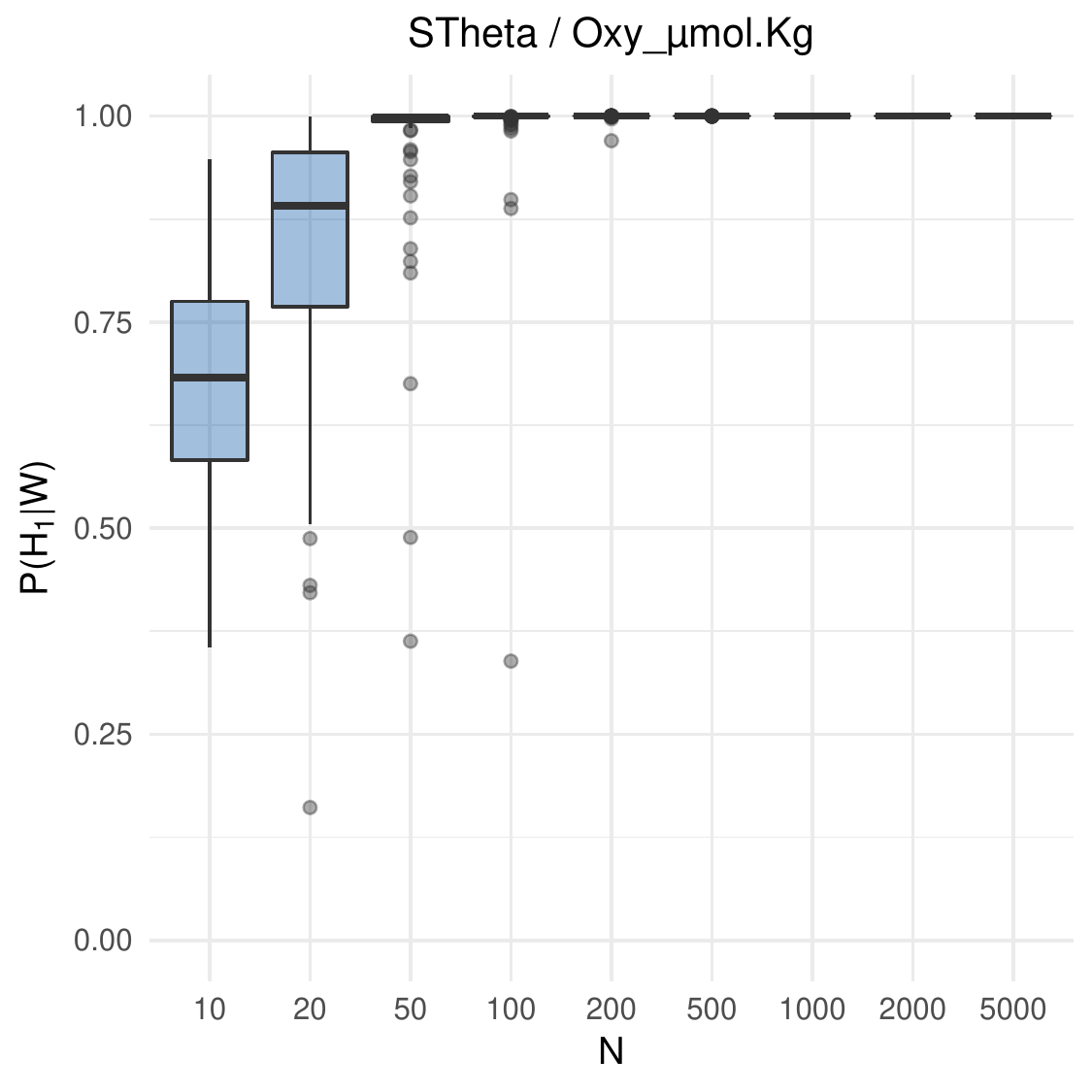}
     \includegraphics[width=0.45\textwidth]{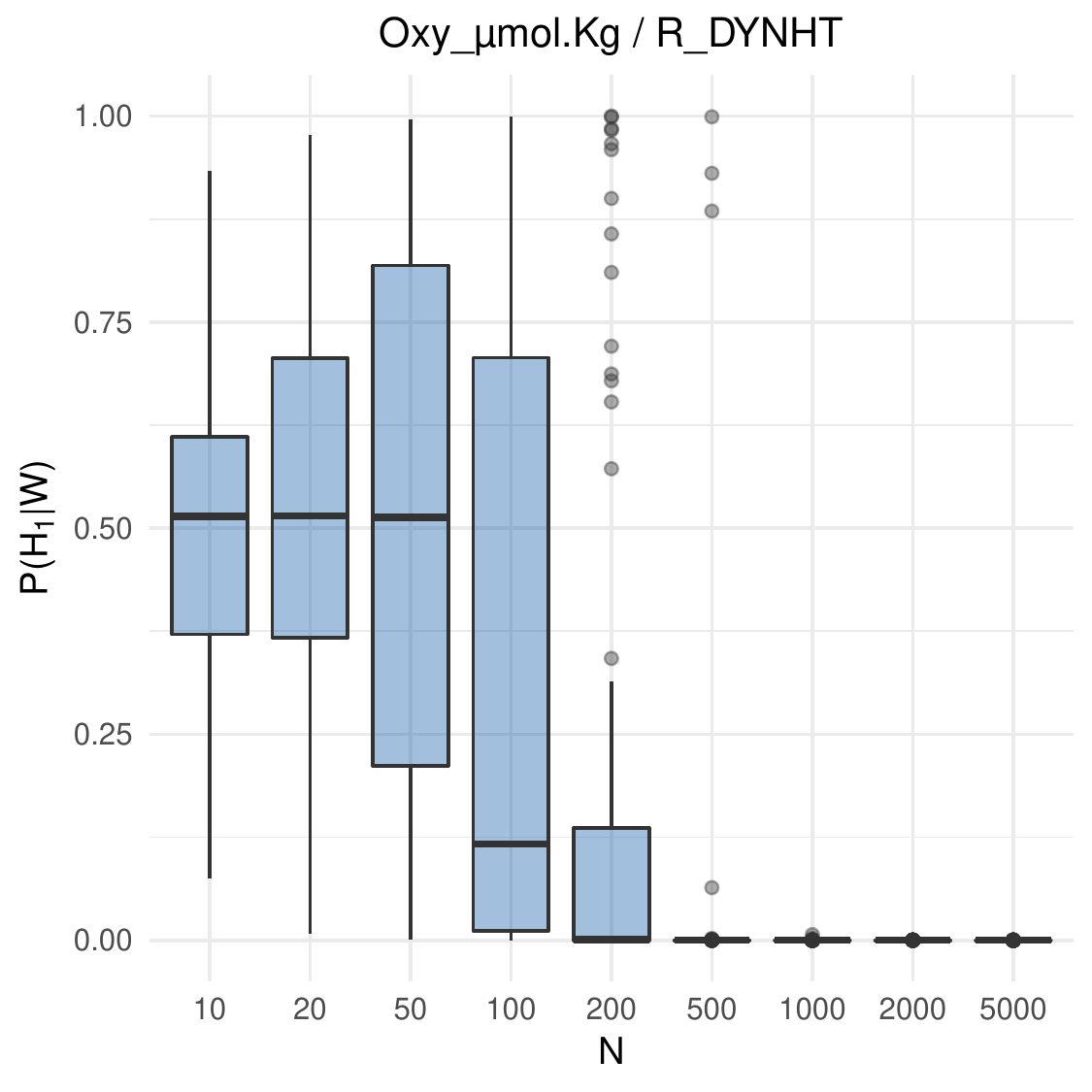}
      \captionsetup{width=.92\linewidth}
      \caption{Box-plots giving the output posterior probability of conditional dependence $p(H_1|W^{(N)})$ for 100 repetitions of the Bayesian conditional independence test applied to randomly-drawn subsamples of various sizes $N$ from the CalCOFI dataset. The left-hand plot gives a representative example of a pair of variables conditionally dependent given \texttt{T\_degC}, while the right-hand plot gives a representative conditionally independent pair.} \label{figure: boxplots}
\end{figure}

\subsection{Real data}
We now apply our method to a representative real dataset to further illustrate its potential. We consider the California Cooperative Oceanic Fisheries Investigations (CalCOFI) Bottle data, a collection of hydrographic readings from maritime stations off the Californian coast collected over a period of 70 years. These data (available at \texttt{calcofi.org}) contain numerous examples of variables with highly non-linear or even non-functional dependence relations. This is illustrated in Figure \ref{figure: calcofi}, which shows pairwise scatter plots of representative data from three of the variables in the dataset.

The complete dataset consists of $864,863$ observations of $74$ variables, but with a high incidence of missing data and numerous strong `trivial' linear correlations. As a consequence, we first remove all variables for which there is at least one other variable with which it has no common data at all. We then calculate pairwise correlation between the remaining variables, and retain only one representative from groups with pair correlations all greater than 0.99. This leaves 657,216 observations of six variables, these being \texttt{T\_degC} (Temperature), \texttt{Salnty} (Salinity), \texttt{STheta} (Potential density), \texttt{Oxy\_µmol.Kg} (Oxygen in micromoles per kg), \texttt{R\_DYNHT} (Dynamic height) and \texttt{R\_PRES} (Pressure). 

For the purposes of exposition, we focus on the case where $Z$ is the variable \texttt{T\_degC}, and $X$ and $Y$ are chosen from the remaining five variables. Though the number of observations remaining even after pre-processing is not significantly lower than in the full dataset, we subsample sets of much smaller cardinality to demonstrate the ability of our test to correctly identify conditional dependence relations in limited data settings. This also serves to effectively eliminate correlation between observations---which would otherwise be strong in this type of time series data---meaning we are able to avoid violating the assumption of i.i.d. data.

For each of the 10 possible pairs $(X,Y)$ chosen from the five remaining variables, we subsample $N=50$, $100$, $200$ then $500$ observations and denote the resulting partial datasets $W^{(N)}$. Figure \ref{figure: graph} gives the graph corresponding to the pairwise dependences found among these five variables conditional on \texttt{T\_degC} for one example draw of each size of subsample, where the number associated with each graph edge is the posterior probability of dependence $p(H_1|W^{(N)})$. The uncertainty in the existence or otherwise of a conditional dependence relation is reflected in the smaller $N$ cases by the posterior probabilities shown for those edges, which are away from 0 and 1. This type of output would be unavailable with a classical test. By $N=200$ (and certainly by $N=500$), the recovered graph emerges clearly. All probabilities are given to two decimal places; where no edge is shown, this indicates $p(H_1|W^{(N)})<0.005$.

The graphs in Figure \ref{figure: graph} are given as an example to show the nature of the output possible with the use of a Bayesian algorithm for this problem. In Figure \ref{figure: boxplots} we give aggregated box-plots for 100 repetitions of the above procedure---analogous to the plots in Figure \ref{table: 4 examples}---for two example variable pairs. These show, as expected, a range in the output posterior probabilities of dependence for the smaller values of $N$. If some sort of thresholding were implemented to produce the equivalent decision output of a classical test (for example: ``reject $H_0$ if $p(H_1|W^{(N)})>0.5$, do not reject $H_0$ otherwise"), then this set of test runs could be used as the basis of an empirical power analysis.

This type of output is in many ways more informative than the output of a classical test. As can be seen in Figure \ref{figure: boxplots}, the algorithm occasionally returns what appears to be a fully incorrect answer (ie. a posterior probability of 1 when the majority of other runs strongly imply a state of conditional independence). This is the equivalent of a classical Type I error. More often, however, the algorithm returns a probability value strictly between 0 and 1---this output is richer and can be interpreted by the analyst more readily as representing an uncertain test outcome.

\subsection{Implementation} \label{sec: hyperparameters}
Practical implementation of the proposed algorithm given in Figure \ref{pseudocode} requires the setting of the two hyperparameters $c$ and $\rho$ as well as recursively-constructed partitions for the various sample spaces. The locations of the splits in such partitions is known to affect inference in the P\'olya tree family of models~\citep{paddock99}. As a default, we suggest two practical approaches to the reader. In the case of a sample space $\Omega$ with compact support, a simple binary partitioning consists of subdividing each set into two subsets of equal size. For $\Omega=[0,1]$, we thus obtain the partition defined by \eqref{binary partition def}.
This is the approach used for the TPT models in the experiments above. Similarly, a quaternary recursive partition of $[0, 1] \times [0, 1]$ can be constructed by subdividing each two-dimensional set into four square quadrants of equal size. 

Another approach, which is also suitable for non-compact sample spaces, is to construct a partition based on the quantiles of a pre-defined distribution $G$; a Gaussian distribution is typically used. For our purposes, it is clear that the partitions of $\Omega_X$, $\Omega_Y$ and $\Omega_Z$ should be constructed separately in order to preserve independence relations. The quaternary recursive partition of $\Omega_{XY}=\Omega_X\times\Omega_Y$ is then constructed from the two binary recursive partitions of $\Omega_X$ and $\Omega_Y$.  The parameters of the distribution $G$---such as the mean and variance in the case that $G$ is Gaussian---can be derived from empirical estimates of the location and \mbox{spread of the samples. }

The mixing parameter $\rho$ controls the probability of stopping during the partitioning of $\Omega_Z$ and thereby defines the balance between the contributions of the restricted-data marginal likelihoods from different scales in the multi-scale mixture model. We have chosen to keep $\rho$ independent of the set $A$, however there is no theoretical impediment to letting $\rho$ depend on $A$. Wong \& Ma \cite{wong10} and Ma \cite{ma17} both fix $\rho = 0.5$ for all their simulations and provide no further discussion of it.

The second hyperparameter is the constant $c$ in the level-dependent Dirichlet hyperparameter $\alpha_j = cj^2$. The question of how to set this is present in all work on P\'olya trees and is in general open. Berger \& Guglielmi \cite{berger01} write that $c$ ``is very difficult to specify'', and it is clear that its value can affect inference. Hanson \& Johnson \cite{hanson02} point out that in the case of the TPT, the limit $c\rightarrow 0$ essentially turns the model into the empirical distribution of the data, while the opposite limit $c\rightarrow \infty$ approaches the parametric model defined by the base measure. In practice, $c=1$ is a common (though ultimately arbitrary) default choice. Other strategies, such as empirical estimation of $c$, have recently been considered \cite{zhang19}. 

\begin{figure}[t]
     \includegraphics[width=0.50\textwidth]{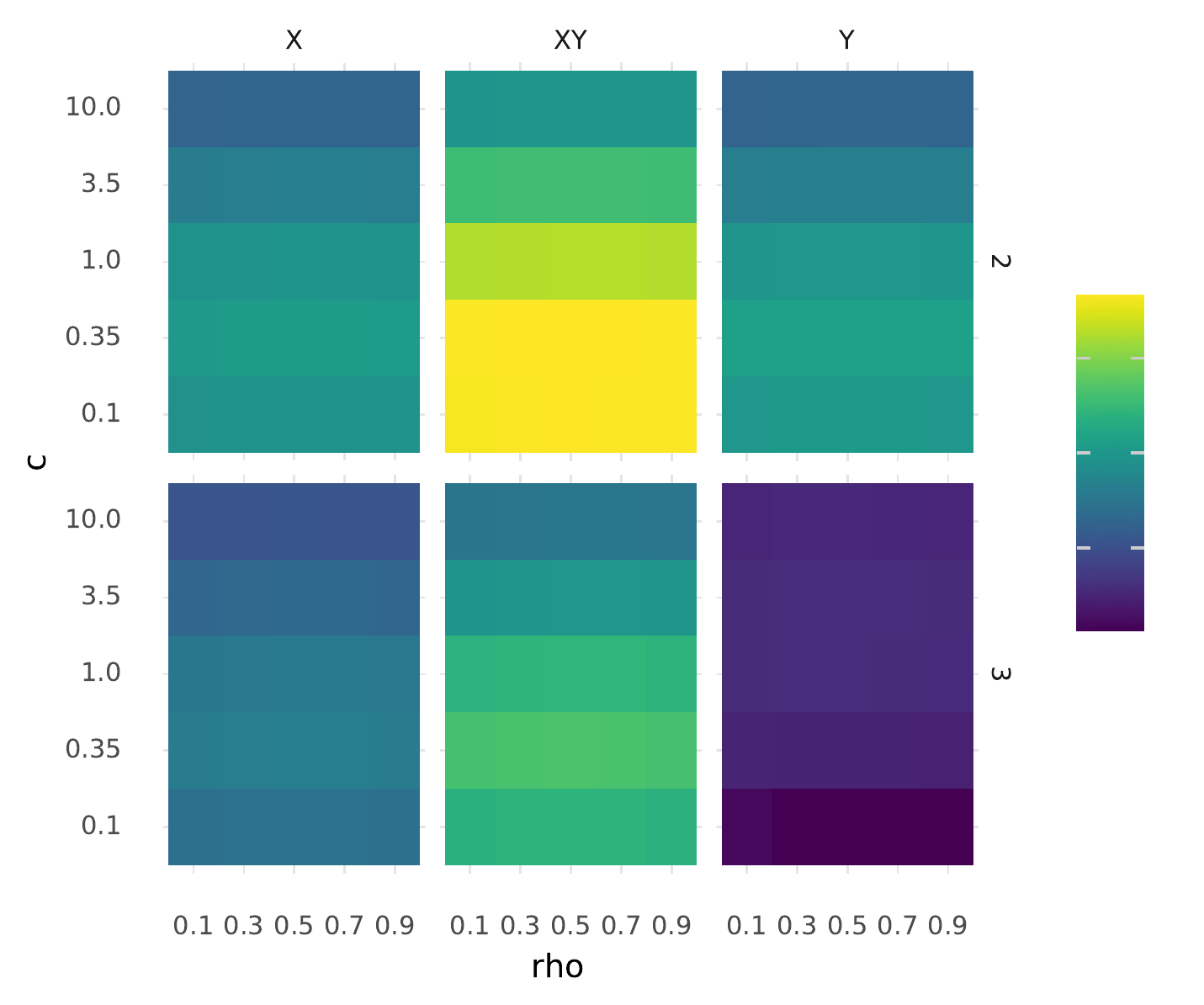}
    \includegraphics[width=0.355\textwidth]{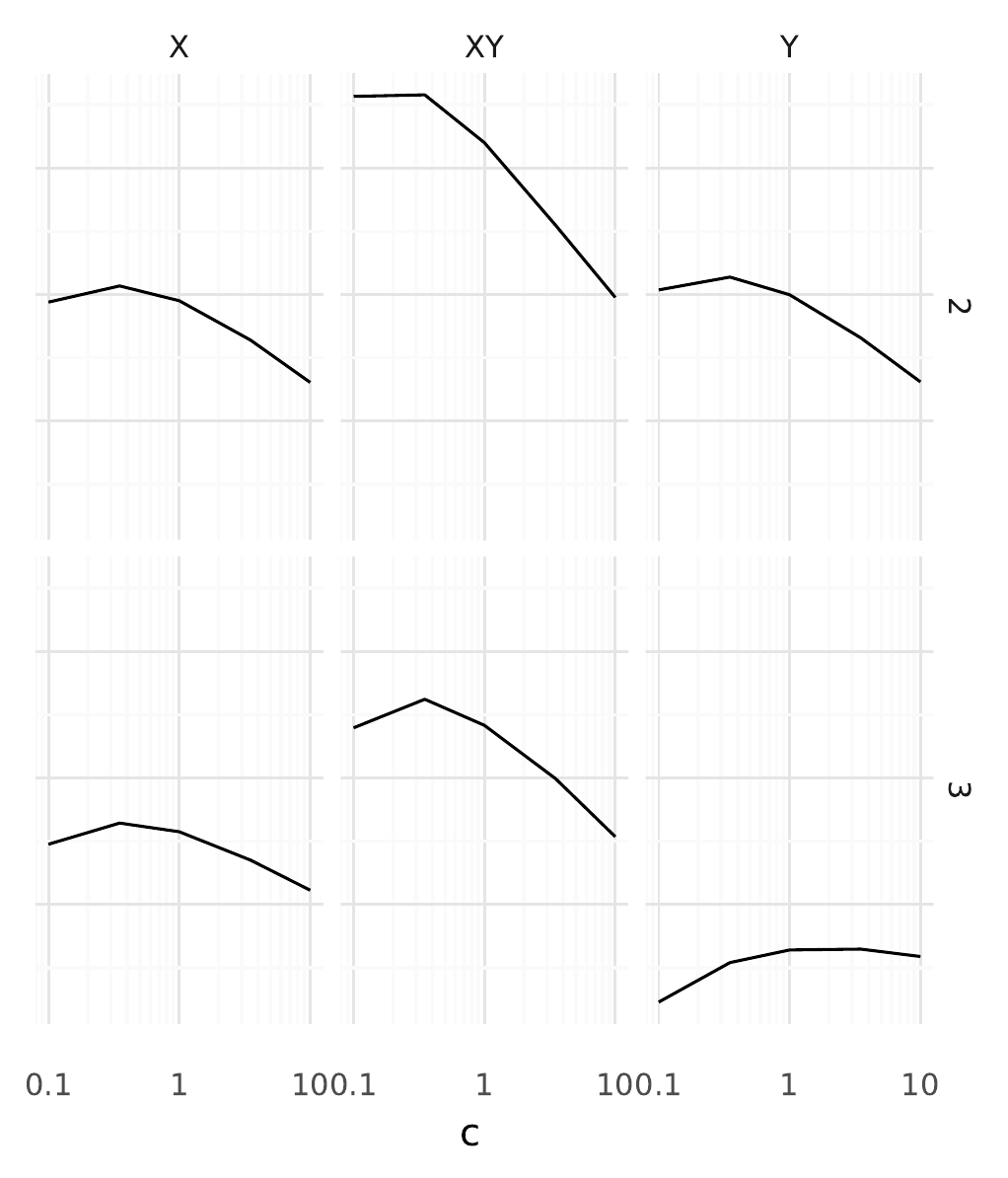}
    \includegraphics[width=0.9\textwidth]{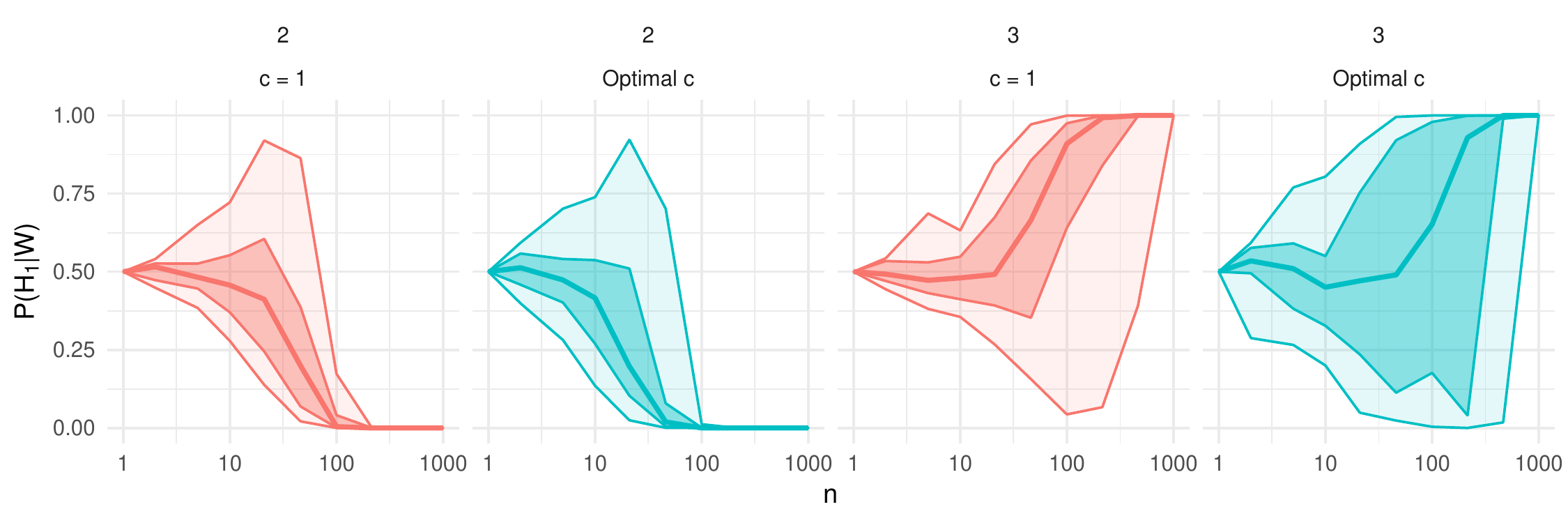}
      \captionsetup{width=.92\linewidth}
      \caption{\emph{Top light:} Heat map of conditional marginal likelihood values for the three constituent models over $\Omega_X$, $\Omega_Y$ and $\Omega_XY$ for the second and third models of Figure \ref{table: 4 examples}. \emph{Top right:} `Slices' from this heatmap with $\rho=0.5$. \emph{Bottom:} Test outputs for 100 repetitions of the second and third models of Figure \ref{table: 4 examples}. Red plots fix $c=1$ (output identical to Figure \ref{table: 4 examples}), while the blue plots use the optimising values $\hat{c}$ from the plot above.} \label{fig: hyperparameter figures}
\end{figure}

We ran preliminary studies to gauge the effect that $\rho$ and $c$ have on the output of our test for the second and third example models in Figure \ref{table: 4 examples}. Plots of marginal likelihood values, with each of the three models over $\Omega_X$, $\Omega_Y$ and $\Omega_{XY}$ considered separately, are given in Figure \ref{fig: hyperparameter figures}. From these it is possible to note the relative lack of sensitivity of the conditional marginal likelihood to variations in $\rho$ for values between approximately $0.3$ and $0.7$. Similar conclusions could be drawn from the remaining two models, not shown here. This evidence, paired with the stated approach of Wong \& Ma \cite{wong10} and Ma \cite{ma17}, justifies our setting $\rho=0.5$ throughout.

As expected, there is a greater degree of sensitivity amongst the individual marginal likelihood values to the value of $c$. In Figure \ref{fig: hyperparameter figures} we contrast the effect on the output posterior probability of conditional independence of setting $c=1$ throughout (as in Figure \ref{table: 4 examples}), and of setting $c$ to the value that maximises the conditional marginal likelihood over a grid of test values for each of the three constituent models separately, ie. $\hat{c}_X \approx \mathrm{argmax}_{c>0}\;p_{X|Z}(W;c)$, and similarly for $\hat{c}_Y$ and $\hat{c}_{XY}$.

The heat map (top left pane of Figure \ref{fig: hyperparameter figures}) gives the conditional marginal likelihood values for the three constituent models over $\Omega_X$, $\Omega_Y$ and $\Omega_XY$ for the second and third models of Figure \ref{table: 4 examples}. The line plots (top right pane) are `slices' from this heatmap which fixes $\rho=0.5$ and seeks to identify the optimal value of $c$. We have left the vertical scale off these plots since we are only interested in maxima rather than the actual values of the conditional marginal likelihood.

The four panes at the bottom of Figure \ref{fig: hyperparameter figures} contrast the test output resulting from the  two different approaches to setting $c$, using the same quantile bands as in Figure \ref{fig: hyperparameter figures}. The red plots (identical to those appearing in Figure \ref{fig: hyperparameter figures}) fix $c=1$, while the blue plots use the optimising values from the plot above. Our empirical findings are that, while the value of $c$ does impact the algorithm output, the consistency of the test procedure does not appear to be affected in the large data limit. Theoretical investigation of this assertion would be a fruitful subject for future research.

A more detailed study of the robustness of derived quantities of P\'olya trees to changes in hyperparameters is beyond the scope of the present work. For practitioners we recommend either a `rule of thumb' approach similar to that we have implemented, possibly with a small number of test runs to calibrate, or a more detailed (but correspondingly more time-consuming) set of pre-simulations. The choice will necessarily be dependent on the dataset under consideration and the balance between speed and accuracy called for by the particular use case.

\section{CONCLUSIONS \& DISCUSSION}

In this article we have defined and demonstrated a new Bayesian nonparametric approach to quantifying the relative evidence in favour of independence or dependence of two random variables conditionally on a third. We have done so in a manner that minimises the assumptions required on the unknown joint distribution of $(X,Y,Z)$, by modelling various of its conditional distributions using P\'{o}lya trees. 

We believe this approach has the potential to be developed in numerous directions, and we hope it will in this way increasingly find application in practical analyses. In its current form, the procedure we describe comes with relatively high computational cost, due primarily to the recursive calculations required, though we hope the line of research opened up by these ideas will soon lead to more efficient implementations. An extension to the multi-dimensional setting, particularly for the conditioning variable $Z$, would be of real use and is the subject of current work.

The Bayesian approach our procedure takes provides a framework in which \emph{both} the hypotheses of conditional independence and conditional dependence can be positively evidenced from a given dataset, unlike the inherently asymmetric hypothesis tests of classical statistics. This is of great importance for causal discovery. 

The output of the procedure is a value in the range $[0,1]$ which can directly be interpreted as a posterior probability of conditional dependence $p(H_1|W)$. This is in notable contrast to previous approaches, even those that work partly within the Bayesian paradigm. This type of output attaches a notion of uncertainty to the result of the test, something absent in classical hypothesis testing. This uncertainty may be propagated further down the `pipeline' of computation if the test is used as a constituent part of a larger procedure.

Our method also allows substantive prior information on the plausibility of an association to be trivially incorporated, something particularly useful when screening large biological datasets. Lastly, the ability to detect dependences of a highly nonlinear or even non-functional nature allows for much greater confidence in the robustness of any inference procedure into which this type of test is embedded.

\newpage
\providecommand{\href}[2]{#2}
\providecommand{\arxiv}[1]{\href{http://arxiv.org/abs/#1}{arXiv:#1}}
\providecommand{\url}[1]{\texttt{#1}}
\providecommand{\urlprefix}{URL }

\medskip
\medskip

\begin{thebibliography}{10}

\bibitem{berger01}
\newblock J.~Berger and A.~Guglielmi,
\newblock Bayesian and conditional frequentist testing of a parametric model
  versus nonparametric alternatives,
\newblock \emph{J. Am. Stat. Assoc.}, \textbf{96}
  (2001), 174--184.

\bibitem{bergsma04}
\newblock W.~Bergsma,
\newblock Testing conditional independence for continuous random variables,
\newblock \emph{Report Eurandom}, 2004.

\bibitem{berrett20}
\newblock T.~B. Berrett, Y.~Wang, R.~F. Barber and R.~J. Samworth,
\newblock The conditional permutation test for independence while controlling
  for confounders,
\newblock \emph{J. R. Stat. Soc. B}, \textbf{82} (2020), 175--197.

\bibitem{candes18}
\newblock E.~Candes, Y.~Fan, L.~Janson and J.~Lv,
\newblock Panning for gold: Model-X knockoffs for high dimensional
  controlled variable selection,
\newblock \emph{J. R. Stat. Soc. B}, \textbf{80} (2018), 551--577.

\bibitem{doran14}
\newblock G.~Doran, K.~Muandet, K.~Zhang and B.~Sch{\"o}lkopf,
\newblock A permutation-based kernel conditional independence test,
\newblock \emph{Proc. 30th Conf. UAI}, 132--141.

\bibitem{escobar95}
\newblock M.~Escobar and M.~West,
\newblock Bayesian density estimation and inference using mixtures,
\newblock \emph{J. Am. Stat. Assoc.}, \textbf{90}
  (1995), 577--588,

\bibitem{filippi17}
\newblock S.~Filippi and C.~Holmes,
\newblock A {{Bayesian}} nonparametric approach to testing for dependence
  between random variables,
\newblock \emph{Bayesian Anal.}, \textbf{12} (2017), 919--938.

\bibitem{fisher24}
\newblock R.~Fisher,
\newblock The distribution of the partial correlation coefficient.,
\newblock \emph{Metron}, \textbf{3} (1924), 329--332.

\bibitem{fukumizu07}
\newblock K.~Fukumizu, A.~Gretton, X.~Sun and B.~Sch\"{o}lkopf,
\newblock Kernel measures of conditional dependence,
\newblock \emph{Adv. Neural Inf. Process. Syst. 20},
  489--496.

\bibitem{ghosal17}
\newblock S.~Ghosal and A.~van~der Vaart,
\newblock \emph{Fundamentals of Nonparametric Bayesian Inference},
\newblock Cambridge University Press, 2017.

\bibitem{ghosh03}
\newblock J.~Ghosh and R.~Ramamoorthi,
\newblock \emph{Bayesian Nonparametrics},
\newblock Springer, 2003.

\bibitem{giudici95}
\newblock P.~Giudici,
\newblock Bayes factors for zero partial covariances,
\newblock \emph{J. Stat. Plan. Inference}, \textbf{46}:2 (1995), 161--174

\bibitem{hanson06}
\newblock T.~Hanson,
\newblock Inference for mixtures of finite {{P\'{o}lya}} tree models,
\newblock \emph{J. Am. Stat. Assoc.}, \textbf{101}
  (2006), 1548--1565.

\bibitem{hanson02}
\newblock T.~Hanson and W.~Johnson,
\newblock Modeling regression error with a mixture of {{P\'{o}lya }}trees,
\newblock \emph{J. Am. Stat. Assoc.}, \textbf{97}
  (2002), 1020--1033.

\bibitem{harris13}
\newblock N.~Harris and M.~Drton,
\newblock {PC} algorithm for nonparanormal graphical models,
\newblock \emph{J. Mach. Learn. Res.}, \textbf{14} (2013),
  3365--3383.
  
\bibitem{hoyer09}
\newblock P.~Hoyer, D.~Janzing, J.~Mooij, J.~Peters and B.~Sch{\"o}lkopf,
\newblock Nonlinear causal discovery with additive noise models,
\newblock \emph{Adv. Neural Inf. Process. Syst. 21},
  \mbox{689--696.}

\bibitem{huang10}
\newblock T.-M. Huang,
\newblock Testing conditional independence using maximal nonlinear conditional
  correlation,
\newblock \emph{Ann. Stat.}, \textbf{38} (2010), 2047--2091,

\bibitem{kass95}
\newblock R.~Kass and A.~Raftery,
\newblock Bayes factors,
\newblock \emph{J. Am. Stat. Assoc.}, \textbf{90}
  (1995), 773--795.

\bibitem{kunihama16}
\newblock T.~Kunihama and D.~B. Dunson,
\newblock Nonparametric {{Bayes}} inference on conditional independence,
\newblock \emph{Biometrika}, \textbf{103} (2016), 35--47.

\bibitem{lavine92a}
\newblock M.~Lavine,
\newblock Some aspects of {{P\'{o}lya}} tree distributions for statistical
  modelling,
\newblock \emph{Ann. Stat.}, \textbf{20} (1992), 1222--1235.

\bibitem{lavine94}
\newblock M.~Lavine,
\newblock More aspects of {{P\'{o}lya}} tree distributions for statistical
  modelling,
\newblock \emph{Ann. Stat.}, \textbf{22} (1994), 1161--1176.

\bibitem{ma13}
\newblock L.~Ma,
\newblock Adaptive testing of conditional association through recursive mixture
  modeling,
\newblock \emph{J. Am. Stat. Assoc.}, \textbf{108}
  (2013), 1493--1505.

\bibitem{ma17}
\newblock L.~Ma,
\newblock Recursive partitioning and multi-scale modeling on conditional
  densities,
\newblock \emph{Electron. J. Stat.}, \textbf{11} (2017),
  1297--1325.

\bibitem{mackay03}
\newblock D.~J. MacKay,
\newblock \emph{Information Theory, Inference and Learning Algorithms},
\newblock Cambridge University Press, 2003.

\bibitem{margaritis05}
\newblock D.~Margaritis,
\newblock Distribution-free learning of bayesian network structure in
  continuous domains,
\newblock \emph{Proc. 20th Nat. Conf. Artificial
  Intel.},
\newblock (2005)
\newblock 825--830.

\bibitem{mauldin92}
\newblock R.~Mauldin, W.~Sudderth and S.~Williams,
\newblock {P\'{o}lya} trees and random distributions,
\newblock \emph{Ann. Stat}, \textbf{20} (1992), 1203--1221.

\bibitem{paddock99}
\newblock S.~Paddock,
\newblock \emph{Randomized {{P\'{o}lya}} trees: {{Bayesian}} nonparametrics for
  multivariate data analysis},
\newblock Ph.{{D}}., Duke University, 1999.

\bibitem{pearl09}
\newblock J.~Pearl,
\newblock \emph{Causality: Models, Reasoning and Inference},
\newblock Cambridge University Press, 2009.

\bibitem{peters17}
\newblock J.~Peters, D.~Janzing and B.~Scholkopf,
\newblock \emph{Elements of Causal Inference: Foundations and Learning
  Algorithms},
\newblock MIT Press, 2017.

\bibitem{peters14a}
\newblock J.~Peters, J.~Mooij, D.~Janzing and B.~Sch{\"o}lkopf,
\newblock Causal discovery with continuous additive noise models,
\newblock \emph{J. Mach. Learn. Res.}, \textbf{15} (2014),
  2009--2053.
  
\bibitem{ramsey14}
\newblock J.~Ramsey,
\newblock A scalable conditional independence test for nonlinear,
  non-{{G}}aussian data,
\newblock \emph{arXiv:1401.5031}.

\bibitem{runge17}
\newblock J.~Runge,
\newblock Conditional independence testing based on a nearest-neighbor
  estimator of conditional mutual information,
\newblock \emph{arXiv:1709.01447}.

\bibitem{saad17}
\newblock F.~Saad and V.~Mansinghka,
\newblock Detecting dependencies in sparse, multivariate databases using
  probabilistic programming and non-parametric Bayes,
\newblock \emph{Proc. Mach. Learn. Res.} \textbf{46} (2017),
\newblock 632--641.

\bibitem{shah18}
\newblock R.~Shah and J.~Peters,
\newblock The hardness of conditional independence testing and the generalised
  covariance measure,
\newblock \emph{arXiv:1804.07203}.

\bibitem{spirtes91}
\newblock P.~Spirtes and C.~Glymour,
\newblock An algorithm for fast recovery of sparse causal graphs,
\newblock \emph{Soc. Sci. Comput. Rev.}, \textbf{9} (1991), 62--72.

\bibitem{strobl18}
\newblock E.~Strobl, K.~Zhang and S.~Visweswaran,
\newblock Approximate kernel-based conditional independence tests for fast
  non-parametric causal discovery,
\newblock \emph{J. Causal Inference}, (2019), 20180017.

\bibitem{su07}
\newblock L.~Su and H.~White,
\newblock {A consistent characteristic function-based test for conditional
  independence},
\newblock \emph{J. Econom.}, \textbf{141} (2007), 807--834.

\bibitem{su08}
\newblock L.~Su and H.~White,
\newblock A nonparametric {Hellinger} metric test for conditional independence,
\newblock \emph{Econom. Theory}, \textbf{24} (2008), 829--864,

\bibitem{wong10}
\newblock W.~H. Wong and L.~Ma,
\newblock Optional {{P{\'o}lya}} tree and {{Bayesian}} inference,
\newblock \emph{Ann. Stat.}, \textbf{38} (2010), 1433--1459.

\bibitem{zhang19}
\newblock J.~Zhang, L.~Yang and X.~Wu,
\newblock P\'{o}lya tree priors and their estimation with multi-group data,
\newblock \emph{Stat. Pap.}, \textbf{60} (2019), 849--875.

\bibitem{zhang12}
\newblock K.~Zhang, J.~Peters, D.~Janzing and B.~Sch{\"o}lkopf,
\newblock Kernel-based conditional independence test and application in causal
  discovery,
\newblock \emph{arXiv:1202.3775}.

\bibitem{zhang17}
\newblock Q.~Zhang, S.~Filippi, S.~Flaxman and D.~Sejdinovic,
\newblock Feature-to-feature regression for a two-step conditional independence
  test,
\newblock \emph{Proc. 33rd Conf. UAI}, 2017.

\end{thebibliography}
\end{document}